\documentclass[
showpacs,
floatfix,
aps,  
prl,
twocolumn,
superscriptaddress,
amssymb,
]{revtex4-1}

\usepackage{times}

\usepackage{wasysym}
\usepackage{amssymb}
\usepackage{latexsym}
\usepackage[dvips]{graphicx}

\usepackage{graphicx}
\usepackage{dcolumn}
\usepackage{amsfonts}
\usepackage{bm}
\usepackage{epsfig}
\usepackage{CJK}

\newcommand{\be}{\begin{equation}}
\newcommand{\ee}{\end{equation}}
\newcommand{\bea}{\begin{eqnarray}}
\newcommand{\eea}{\end{eqnarray}}

\def\m{m^\star}

\def\oc{\omega_{\mbox{\scriptsize {c}}}}

\def\rc{R_{\mbox{\scriptsize {c}}}}

\def\tq{\tau_{\rm q}}

\def\ttr{\tau}

\newcommand{\req}[1]{Eq.\,(\ref{#1})}

\newcommand{\reqs}[2]{Eqs.\,(\ref{#1}),\,(\ref{#2})}
\newcommand{\rEqs}[2]{Equations\,(\ref{#1}),\,(\ref{#2})}
\newcommand{\rfig}[1]{Fig.\,\ref{#1}}

\newcommand{\rref}[1]{Ref.\,\onlinecite{#1}}
\newcommand{\rrefs}[2]{Refs.\,\onlinecite{#1},\,\onlinecite{#2}}

\def\ne{n_e}

\def\rxx{R_{xx}}
\def\ryy{R_{yy}}

\def\easy{\left < 110 \right >}
\def\hard{\left < 1\bar10 \right >}
\def\x{\hat{x}}
\def\y{\hat{y}}

\def\ne{n_e}

\def\nul{\nu_1}
\def\nuh{\nu_2}

\begin{document}
\begin{CJK*}{UTF8}{gbsn}
\title{Hidden Quantum Hall Stripes in Al$_{x}$Ga$_{1-x}$As/Al$_{0.24}$Ga$_{0.76}$As Quantum Wells}
\author{X. Fu}
\affiliation{School of Physics and Astronomy, University of Minnesota, Minneapolis, Minnesota 55455, USA}
\author{Yi Huang (黄奕)}
\affiliation{School of Physics and Astronomy, University of Minnesota, Minneapolis, Minnesota 55455, USA}
\author{Q. Shi}
\altaffiliation[Present address: ]{Department of Physics, Columbia University, New York, NY, USA}
\author{B.\,I. Shklovskii}
\affiliation{School of Physics and Astronomy, University of Minnesota, Minneapolis, Minnesota 55455, USA}
\author{M.\,A. Zudov}
\email[Corresponding author: ]{zudov001@umn.edu}
\affiliation{School of Physics and Astronomy, University of Minnesota, Minneapolis, Minnesota 55455, USA}
\author{G.\,C. Gardner}
\affiliation{Microsoft Quantum Lab Purdue, Purdue University, West Lafayette, Indiana 47907, USA}
\affiliation{Birck Nanotechnology Center, Purdue University, West Lafayette, Indiana 47907, USA}
\author{M.\,J. Manfra}
\affiliation{Microsoft Quantum Lab Purdue, Purdue University, West Lafayette, Indiana 47907, USA}
\affiliation{Birck Nanotechnology Center, Purdue University, West Lafayette, Indiana 47907, USA}
\affiliation{Department of Physics and Astronomy, Purdue University, West Lafayette, Indiana 47907, USA}
\affiliation{School of Electrical and Computer Engineering and School of Materials Engineering, Purdue University, West Lafayette, Indiana 47907, USA}
\received{\today}
\begin{abstract}
We report on transport signatures of hidden quantum Hall stripe (hQHS) phases in high ($N > 2$) half-filled Landau levels of Al$_{x}$Ga$_{1-x}$As/Al$_{0.24}$Ga$_{0.76}$As quantum wells with varying Al mole fraction $x < 10^{-3}$.
Residing between the conventional stripe phases (lower $N$) and the isotropic liquid phases (higher $N$), where resistivity decreases as $1/N$, these hQHS phases exhibit isotropic and $N$-independent resistivity.
Using the experimental phase diagram we establish that the stripe phases are more robust than theoretically predicted, calling for improved theoretical treatment. 
We also show that, unlike conventional stripe phases, the hQHS phases do not occur in ultrahigh mobility GaAs quantum wells, but are likely to be found in other systems.
\end{abstract}
\maketitle
\end{CJK*}

Discovery of the integer quantum Hall effect in Si \citep{klitzing:1980} has paved the way to observations of many exotic phenomena in two-dimensional (2D) electron and hole systems.
Two prime examples are the fractional quantum Hall effect \citep{tsui:1982} and quantum Hall stripes (QHSs) \cite{koulakov:1996,fogler:1996,moessner:1996,lilly:1999a,du:1999}.
While fractional quantum Hall effects have been realized in many systems, including GaAs \citep{tsui:1982}, Si \citep{nelson:1992,kott:2014}, AlAs \citep{poortere:2002}, GaN \citep{manfra:2002}, graphene \citep{du:2009,bolotin:2009}, CdTe \citep{piot:2010}, ZnO \citep{tsukazaki:2010}, Ge \citep{shi:2015c}, and InAs \citep{ma:2017}, exploration of the QHS physics remains limited to GaAs \citep{note:nem}.

Forming due to a peculiar boxlike screened Coulomb potential, QHSs can be viewed as charge density waves consisting of stripes with alternating integer filling factors $\nu$, e.g., $\nu = 4$ and $\nu = 5$ \citep{note:fradkin}.
In experiments, QHSs are manifested by giant resistivity anisotropies ($\rho_{xx} \gg \rho_{yy}$) in $N \ge 2$ half-filled Landau levels.
Appearance of these anisotropies in macroscopic samples is attributed to a mysterious symmetry-breaking field \cite{fil:2000,koduvayur:2011,sodemann:2013,pollanen:2015}, which nearly always aligns QHSs along the $\y \equiv \easy$ crystal axis of GaAs \citep{note:ex}.
While a sufficiently low disorder is necessary for the QHS formation, the absence of QHSs in systems beyond GaAs might simply be due to the lack of symmetry-breaking fields \citep{note:ts}.
Indeed, electron bubble phases \cite{koulakov:1996,fogler:1996,moessner:1996,cooper:1999,eisenstein:2002,lewis:2002,deng:2012a,rossokhaty:2016,friess:2017,bennaceur:2018,friess:2018,fu:2019,ro:2019}, which are close relatives of QHSs, have already been identified in graphene \citep{chen:2019}.

In this Letter, we report observation of transport signatures of the recently predicted \citep{huang:2020} hidden QHS (hQHS) phases in a series of Al$_{x}$Ga$_{1-x}$As/Al$_{0.24}$Ga$_{0.76}$As quantum wells with $x < 10^{-3}$.
In contrast to the ordinary QHS phases, the hQHS phases are characterized by isotropic resistivity ($\rho_{xx} = \rho_{yy} = \rho$) that is independent of $\nu$, unlike the isotropic liquid phases in which $\rho \propto \nu^{-1}$.
These unique properties make these phases detectable without symmetry-breaking fields, thereby opening an avenue to study stripe physics in systems beyond GaAs.
The wide variation of mobilities in our samples allows us to construct an experimental phase diagram in the conductivity-filling factor plane. 
Its comparison to theoretical predictions \citep{huang:2020} yields the electron quantum lifetimes and the stripe density of states.
The latter turns out to be lower than predicted by original the Hartree-Fock theory \citep{koulakov:1996,fogler:1996}, calling for further theoretical input.
We confirm this finding by a complementary experiment on an ultrahigh mobility GaAs quantum well, where we also show that,  in this sample, the hQHS phase yields to the QHS phase in agreement with the theory.

Before presenting our experimental data, we briefly summarize the physics behind the hQHS phases \citep{huang:2020}.
The resistance anisotropies in the ordinary QHS phase emerge due to different diffusion mechanisms along and perpendicular to the stripes \citep{macdonald:2000,oppen:2000,sammon:2019}.
In this picture, an electron drifts a distance $L_y$ along the $y$-oriented stripe edge in an $x$-directed internal electric field until it is scattered by impurities to one of the adjacent stripe edges located at a distance $L_x = \Lambda/2 \approx \sqrt{2}\rc$ \citep{koulakov:1996,fogler:1996}, where $\Lambda$ is the stripe period and $\rc$ is the cyclotron radius.
When $L_y \gg L_x$, the diffusion coefficient in the $\hat y$ direction is much larger that in the $\hat x$ direction, which leads to anisotropic conductivity, $\sigma_{yy} \gg \sigma_{xx}$, and resistivity, $\rho_{xx} \gg \rho_{yy}$. 
Since $L_y \propto \nu^{-1}$ and $L_x \propto \nu$ \citep{sammon:2019}, the anisotropy decreases with $\nu$ and eventually vanishes at some $\nu = \nul$.
At larger $\nu$, the drift contribution to the diffusion along stripes can be neglected, and $L_y$, like $L_x$, is determined entirely by the impurity scattering.
For isotropic scattering, it is easy to show \citep{note:ly} that $L_y = \sqrt{2}\rc$ which coincides with $L_x$.
As a result, the QHS phase yields to the hQHS phase in which the resistivity is isotropic and $\nu$ independent (since the stripe density of states does not vary with $\nu$).
The hQHS phase persists until the stripe structure is destroyed by disorder at $\nu = \nuh$ and the ground state becomes an isotropic liquid with $\rho_{xx} = \rho_{yy}\propto \nu^{-1}$, as predicted by Ando and Uemura \citep{ando:1974.a} and experimentally confirmed by Coleridge, Zawadski, and Sachrajda (CZS) \citep{coleridge:1994}.

For the hQHS phase to exist and be detected, it should span a sizable range of the filling factors $\Delta \nu = \nuh - \max\{\nul, 9/2\} \gg 1$.
The range $\Delta \nu$ depends sensitively on both transport $\ttr^{-1}$ and quantum $\tq^{-1}$ scattering rates, which control $\nul$ and $\nuh$, respectively \citep{huang:2020}.
As we will see, ultrahigh mobility GaAs quantum wells do not support the hQHS phase as $\nul \approx \nuh$ in these samples.
On the other hand, adding the correct small amount of Al \citep{note:al} to the GaAs well greatly expands $\Delta \nu$, as it affects $\nul$ to a much greater extent than it does $\nuh$.
This happens because Al acts as a short-range disorder, which contributes \emph{equally} to transport $\ttr^{-1}$ and quantum $\tq^{-1}$ scattering rates, and because $\tq/\ttr \ll 1$ at $x = 0$.

\begin{table}[b]
\vspace{-0.25 in}
\caption{
Sample ID, Al mole fraction $x$, electron density $n_e$, mobility $\mu$, and Drude conductivity, in units of $e^2/h$, $\tilde\sigma_0 = h\ne \mu/e$ at zero magnetic field ($B=0$).
}
\begin{ruledtabular}
\begin{tabular}{c c c c c c} 
Sample ID & $x$  & $n_e$ ($10^{11}$ cm$^{-2}$)  & $\mu$ ($10^6$ cm$^2$/Vs) &$\tilde\sigma_0$ ($10^3$)\\
\hline \\ [-2ex] 
A & 0.00057 & 3.0 &  6.5 & 8.0  \\ 
B & 0.00082 & 2.9 &  4.1 & 4.9  \\ 
C & 0.0078 & 2.7 &  1.2 &  1.3   
\end{tabular}
\end{ruledtabular}
\label{table:a}
\end{table}

Apart from different $x$, all our Al$_{x}$Ga$_{1-x}$As quantum wells share an identical heterostructure design \cite{gardner:2013}.
Electrons are supplied by Si doping in narrow GaAs wells surrounded by narrow AlAs layers and placed at a setback distance of $75$ nm from each side of the 30-nm-wide Al$_{x}$Ga$_{1-x}$As well hosting the 2D electrons.
Parameters of samples A, B, and C such as Al mole fraction $x$, electron density $n_e$, mobility $\mu$, and Drude conductivity $\tilde\sigma_0 = h\ne \mu/e$ in units of $e^2/h$ at zero magnetic field ($B=0$) are listed in Table\,\ref{table:a}.
The samples are approximately 4 mm squares with eight indium contacts positioned at the corners and at the midsides.
Longitudinal resistances $R_{xx}$ and $R_{yy}$ were measured in sweeping magnetic fields using a four-terminal, low-frequency (a few Hz) lock-in technique at a temperature $T \approx 25$ mK at which the resistances are nearly temperature independent.
The current was sent along either the $\x \equiv \hard$ or $\y \equiv \easy$ direction using the midside contacts, and the voltage was measured between contacts along the edge.
To account for anisotropies due to nonideal geometry, $\rxx$ or $\ryy$ was multiplied by a factor (typically $\lesssim 1.1$) that was chosen to make $\rxx = \ryy$ in the low field regime. 

\begin{figure}[t]
\includegraphics{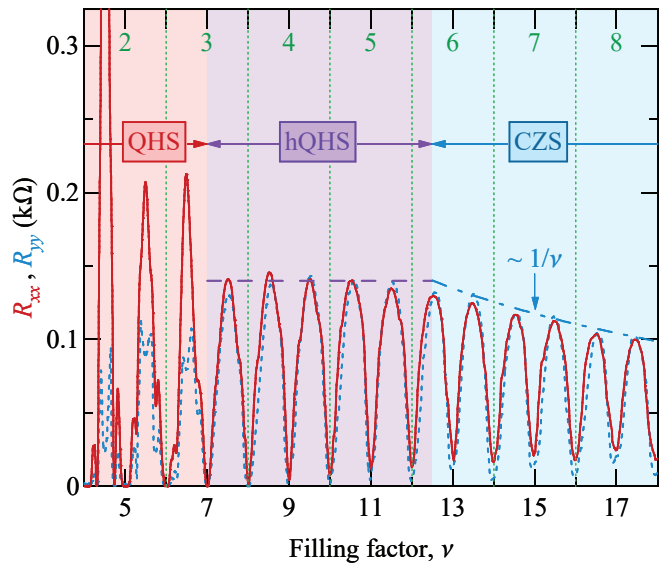}
\vspace{-0.1 in}
\caption{
Longitudinal resistances $\rxx$ (solid line) and $\ryy$ (dotted line) as a function of the filling factor $\nu$ measured in sample B.
Gap centers between spin-resolved Landau levels are labeled by $N =2,3,...$, at the top axis ($\nu = 2N + 1$).
The conventional QHS phase ($\rxx > \ryy$) and the CZS phase ($\rxx \approx \ryy \propto \nu^{-1}$) occur at half-integer $\nu = 9/2, 11/2, 13/2$ and at $\nu = 27/2, 29/2,...$, respectively.
The hQHS phase is identified at intermediate half-integer filling factors, $\nu = 15/2,...., 25/2$, where the resistance is isotropic \emph{and} $\nu$ independent.
The characteristic $\nu^{0}$ ($\nu^{-1}$) dependence of the isotropic resistance in the hQHS (CZS) phase is marked by dashed (dash-dotted) line.
 }
\label{fig:data}
\vspace{-0.15 in}
\end{figure}

In \rfig{fig:data}, we present longitudinal resistances $\rxx$ and $\ryy$ as a function of filling factor $\nu$ measured in sample B.
At low half-integer filling factors ($\nu = 9/2$, $11/2$, and $13/2$) the data reveal conventional QHS phases, as evidenced by $\rxx > \ryy$.
At high half-integer filling factors ($\nu > 25/2$) we identify the CZS phase in which $\rxx \approx \ryy \propto \nu^{-1}$ (cf. dash-dotted line).
At intermediate half-integer filling factors, $\nu = 15/2,..., 23/2$, one readily confirms {\it both} characteristic features of the hQHS phase; indeed, the data show that two longitudinal resistances are practically the same ($\rxx \approx \ryy$) and are \emph{independent} of $\nu$ (cf. dashed line).
From \rfig{fig:data}, we can easily identify the characteristic filling factors $\nu_1 \approx 7$ and $\nu_2 \approx 12.5$ which mark the crossovers from the QHS to the hQHS phase and from the hQHS to the CZS phase, respectively.

\begin{figure}[t]

\includegraphics{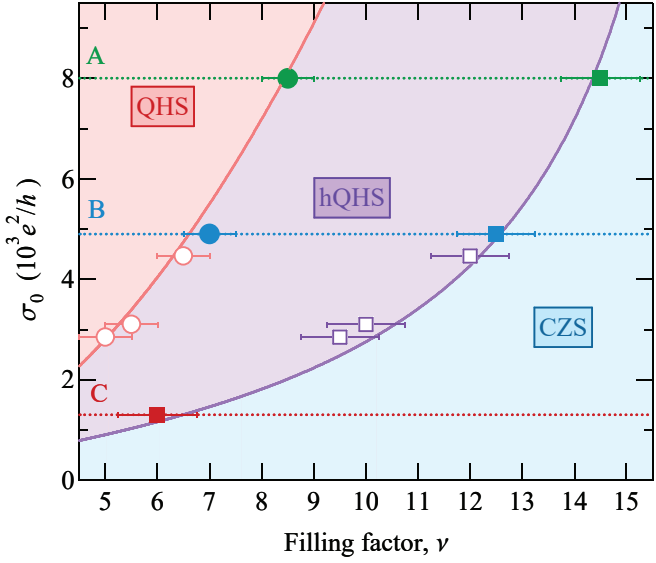}
\vspace{-0.1 in}
\caption{A diagram in the $(\nu,\sigma_0)$ plane showing QHS, hQHS, and CZS phases.
Solid lines represent crossovers between phases, \req{eq:nul} [left(upper) line] and \req{eq:nuh} [right(lower) line]. 
Solid circles (solid squares) represent experimental $\nul$ ($\nuh$) and horizontal dotted lines mark $\tilde \sigma_0$ for samples A-C \citep{gardner:2013}.
Open circles (squares) are additional data from a study conducted in a different context that conform to our present findings \citep{shi:thesis}.
}
\label{fig:diag}
\vspace{-0.15 in}
\end{figure}

In a similar manner, we have obtained $\nul$ and $\nuh$ for sample A and $\nuh$ for sample C (which does not support the QHS phase due to higher Al mole fraction $x$), which we then use to construct the experimental phase diagram shown in \rfig{fig:diag}.
We start by adding points representing the dimensionless Drude conductivity $\tilde \sigma_0$ for samples A, B, C (see Table\,\ref{table:a}) and the corresponding filling factors $\nul$ (solid circles) and $\nuh$ (solid squares).
To connect these data points we use the theoretical boundaries of the hQHS phase \citep{huang:2020}.
The lower boundary $\nu = \nul$, separating the QHS and the hQHS phases, is given by \citep{huang:2020}
\be
\nul \simeq  \frac{\sqrt{\tilde \sigma_0}} {\alpha}\,,
\label{eq:nul}
\ee 
where $\alpha$ \citep{note:gamma} is the QHS density of states in units of the density of states per spin at $B = 0$, $g_0 = \m/2\pi \hbar^2$.
This boundary can be obtained by either matching the parameter-free geometric average of the resistivities in the QHS phase $\sqrt{\rho_{xx}\rho_{yy}} = (h/e^2)/(2\nu^2+1/2)\approx (h/e^2)/2\nu^2$ \citep{macdonald:2000,sammon:2019} and the resistivity in the hQHS phase \citep{huang:2020},
\be
\tilde \rho_{\rm hQHS} \equiv \frac h {e^2} \frac{\alpha^2}{2\tilde\sigma_0}\,,
\label{eq:hqhs}
\ee
or, equivalently, by setting the resistivity anisotropy ratio to unity, $\rho_{xx}/\rho_{yy} \approx (\tilde \sigma_0/\alpha^2\nu^2)^2 = 1$ \citep{sammon:2019,huang:2020}.

The higher boundary $\nu =\nuh$ marks the crossover from the hQHS to the CZS phase and is represented by
\be
\nuh 
\simeq \frac {\tilde \sigma_0} {\alpha^2} \frac \tq \ttr\,.
\label{eq:nuh}
\ee
This boundary can be obtained by equating $\alpha$ and the density of states at the center of the Landau level in CZS phase, in units of the density of states at $B = 0$, $\sqrt{\oc\tq}$ \citep{raikh:1993,mirlin:1996} or by matching $\rho_{\rm hQHS}$ and the resistivity in the CZS phase \citep{coleridge:1994}, 
\be
\rho_{\rm CZS} \equiv \frac h {e^2} \frac 1 \nu \frac {\tq/2\tau}{(\tq/2\tau)^2+1} \approx \frac h {e^2} \frac 1 \nu \frac {\tq}{2\tau}\,.
\label{eq:czs}
\ee
We thus see that for a given carrier density, as mentioned above, $\nuh$ and $\nul$ are controlled by $\tau$ and $\tq$, respectively.
Strictly speaking, \reqs{eq:nul}{eq:nuh} are not sharp boundaries but rather gradual crossovers between corresponding phases.

With the help of \req{eq:nul} and experimental values of $\nul$ in samples A and B, we estimate $\alpha \approx 11$, which is smaller than the theoretical estimate of $\alpha \simeq 18$ \citep{sammon:2019,note:gamma}.
We then parameterize scattering rates $\ttr^{-1}$ and $\tq^{-1}$ as
\be
\tau^{-1} = \tau_0^{-1} + \kappa x\,,~~~\tq^{-1} = \tau_{\rm q0}^{-1} + \kappa x\,,
\label{eq:tau}
\ee
where $x$ is the Al mole fraction, $\kappa \approx 24$ ns$^{-1}$ per \% Al \citep{gardner:2013}, and $\tau_0^{-1} \approx 3$ ns$^{-1}$ \citep{gardner:2013} is the transport scattering rate in the limit of $x \rightarrow 0$.
To find the remaining parameter $\tau_{\rm q0}^{-1}$, which is the quantum scattering rate in the limit of $x \rightarrow 0$, we use experimental $\nu_2$ values and notice that \reqs{eq:nul}{eq:nuh} yield $\tq/\ttr \simeq \nuh/\nul^2$.
Using \req{eq:tau} we then obtain an estimate for $\tau_{\rm q0} \simeq 0.05$ ns which is in good agreement with $\tq$ values found from low $B$ experiments \citep{shi:2016a,shi:2017a,zudov:2017} on microwave-induced \citep{zudov:2001a,ye:2001, dmitriev:2012} and Hall-field-induced \citep{yang:2002,zhang:2007a,vavilov:2007} resistance oscillations in GaAs quantum wells.

We next use $\ne = 3 \times 10^{11}$ cm$^{-2}$ and $\m  = 0.06\,m_0$ \citep{coleridge:1996,tan:2005,hatke:2013,shchepetilnikov:2017,fu:2017} to compute the phase boundaries, \reqs{eq:nul}{eq:nuh}, which are shown in \rfig{fig:diag} by solid lines.
Both lines pass in close proximity to the experimentally obtained $\nu_1$ (solid circles) and $\nu_2$ (solid squares) from all samples, showing excellent agreement between theory \citep{huang:2020} and experiment.
Finally, we add data points (open circles and squares) from three other Al$_{x}$Ga$_{1-x}$As/Al$_{0.24}$Ga$_{0.76}$As quantum wells that were investigated in a different context \citep{shi:thesis}. 
These points are also in agreement with the theory and the present experiment.

\begin{figure}[t]
\includegraphics{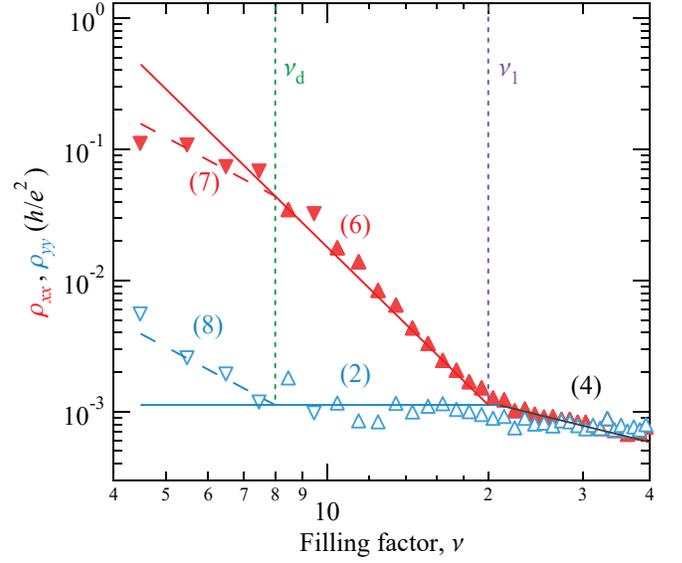}
\vspace{-0.1 in}
\caption{
$\rho_{xx}$ (solid triangles) and $\rho_{yy}$ (open triangles) \citep{note:symbols}) as a function of filling factor $\nu$ for sample A of \rref{sammon:2019}.
Lines are computed using theoretical expressions, marked by equation numbers.
}
\label{fig:clean}
\vspace{-0.15 in}
\end{figure}

Having confirmed the existence of the hQHS phases in Al$_{x}$Ga$_{1-x}$As/Al$_{0.24}$Ga$_{0.76}$As quantum wells, we next examine the possibility for these phases to exist in ultrahigh mobility GaAs quantum wells (without alloy disorder).
In such samples, the lower boundary $\nul$, \req{eq:nul}, might approach and even merge with the higher boundary $\nuh$, \req{eq:nuh}, eliminating the hQHS phase as a result.
To test this scenario, we revisit the data obtained from sample A of \rref{sammon:2019} with $\tilde\sigma_0 \approx 3.4 \times 10^4$, much higher than in samples used in the present study.
As illustrated in \rfig{fig:clean}, showing $\rho_{xx}$ (solid triangles) and $\rho_{yy}$ (open triangles) \citep{note:symbols} as a function of the filling factor $\nu$, the QHS anisotropy in this sample collapses at $\nul \approx 20$.
Using \req{eq:nul}, we can then estimate $\alpha  = \sqrt{\tilde \sigma_0}/\nul \approx 9$ \citep{note:17}.
With $\tq \simeq 0.05$ ns, \req{eq:nuh} gives $\nuh \approx 21$, which is very close to $\nul \approx 20$.
Indeed, the data in \rfig{fig:clean} show that the QHS phase crosses over directly to the CZS phase, bypassing the intermediate hQHS phase.

In the QHS phase, the easy resistivity is $\nu$ independent and is described by $\rho_{yy} = \rho_{\rm hQHS}$, \req{eq:hqhs}, while the hard resisitivty exhibits clear $\nu^{-4}$ dependence and follows \citep{huang:2020} 
\be
\rho_{xx} \simeq \frac h {e^2} \frac {\tilde \sigma_0} {2\alpha^2 \nu^4}\,.
\label{eq:rhoxx}
\ee
However, the agreement between theory and experiment breaks down at $\nu < \nu_{\rm d} \approx 8$, where one observes significant deviations leading to the reduction of the anisotropy.
While the nature of such reduction is unclear, it becomes more pronounced upon further cooling and might reflect a crossover to another competing ground state \citep{qian:2017,fu:2020a}.
We can account for the observed anisotropy reduction at lower filling factors assuming that the QHS phase has a finite concentration of dislocations separated by an average distance $L_{\rm d} = \beta \Lambda/2$ along stripes, where $\beta$ is a numerical factor. 
Scattering of drifting electrons by these dislocations limits their drift length by $L_{\rm d} \ll L_y$ and the resistivities calculated in \rrefs{sammon:2019}{huang:2020} need to be modified to \citep{note:dis,note:dis1}
\be
\rho_{xx} = \frac h {e^2} \frac{\beta}{2\nu^2}\,,
\label{eq:rhoxx_dis}
\ee
\be
\rho_{yy} = \frac h {e^2} \frac{1}{2 \beta \nu^2}\,.
 \label{eq:rhoyy_dis}
\ee
\rEqs{eq:rhoxx_dis}{eq:rhoyy_dis} are plotted as dashed lines in \rfig{fig:clean}. 
Equating \req{eq:rhoxx_dis} to \req{eq:rhoxx} [or \req{eq:rhoyy_dis} to \req{eq:hqhs}], we find that the crossover to the dislocation limited transport happens at
\begin{equation}
\nu_{\rm d} \equiv \frac{\nul} {\sqrt{\beta}}\,.
\end{equation}
With $\nu_{\rm d} \simeq 8$ and $\nul \simeq 20$ we find $\beta  = (\nul/\nu_{\rm d})^2 \simeq 6.3$. 
This value does not seem unreasonable and correctly accounts for the saturation of the anisotropy, $\rho_{xx}/\rho_{yy} =\beta^2 \approx 40$ \citep{sammon:2019}.

\begin{figure}[t]
\includegraphics{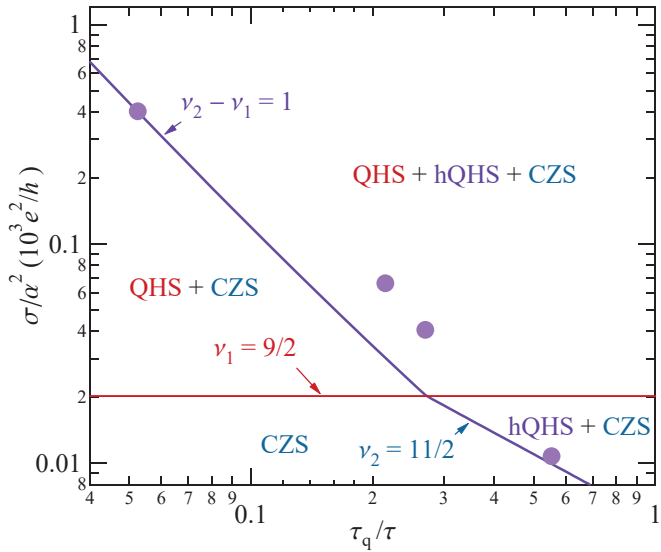}
\vspace{-0.1 in}
\caption{A diagram in the $(\tq/\ttr,\sigma_0/\alpha^2)$ plane showing four regions marked by detectable phases.
Circles are experimental data points from all four samples studied.
}
\label{fig:diag2}
\vspace{-0.15 in}
\end{figure}

Our experimental findings in Al$_x$Ga$_{1-x}$As quantum wells (\rfig{fig:diag}) and in a clean GaAs quantum well (\rfig{fig:clean}) can be unified in a phase diagram shown in \rfig{fig:diag2} which treats $\sigma_0/\alpha^2$ and $\tq/\ttr$ as independent parameters.
Here, the QHS phase is observed above the horizontal line corresponding to $\nu_1 = 9/2$.
To detect the hQHS phase, one should satisfy both $\nuh - \nul > 1$  and $\nuh > 11/2$, since at least two half-integer filling factors are needed to establish the $\nu$ independence of the resistance \citep{note:b}.
As a result, the most favorable conditions for the hQHS phase are realized at the top-right corner of the diagram.
However, as demonstrated by our experiments on Al$_x$Ga$_{1-x}$As quantum wells, the hQHS can be detected at modest mobilities provided that the ratio $\tq/\ttr$ is sufficiently high.
On the other hand, this ratio is much smaller in clean GaAs quantum wells, which makes the hQHS detection difficult in such systems despite their high mobility.
The phase diagram shown in \rfig{fig:diag2} provides a a road map for future experiments aiming to detect the hQHS phases.

In summary, we have observed hidden quantum Hall stripe (hQHS) phases \citep{huang:2020} forming near half-integer filling factors of Al$_{x}$Ga$_{1-x}$As/Al$_{0.24}$Ga$_{0.76}$As quantum wells with varying $x$. 
These phases reside between the conventional stripe phases and the isotropic liquid phases and are characterized by isotropic resistivity that is not sensitive to the filling factor.
Analysis of the experimental phase diagram reveals that the QHS density of states is smaller than predicted by the Hartree-Fock theory \citep{koulakov:1996,fogler:1996}, calling for improved theory.
The unique transport characteristics of the hQHS phases should allow exploration of the stripe physics in 2D systems that, unlike GaAs, lack symmetry-breaking fields. 
On the other hand, ultrahigh mobility GaAs quantum wells favor conventional QHSs over hQHSs due to a shrinking filling factor range where the hQHS phases can form.

\begin{acknowledgments}
We thank G. Jones, T. Murphy, and A. Bangura for technical support.
Calculations by Y.H. were supported primarily by the National Science Foundation through the University of Minnesota MRSEC under Award Number Nos. DMR-1420013 and DMR-2011401.
Experiments by X.F., Q.S., and M.Z. were supported by the U.S. Department of Energy, Office of Science, Basic Energy Sciences, under Award DE-SC0002567.
Growth of quantum wells at Purdue University was supported by the U.S. Department of Energy, Office of Science, Basic Energy Sciences, under Award DE-SC0006671.
X.F. acknowledges the University of Minnesota Doctoral Dissertation Fellowship.
A portion of this work was performed at the National High Magnetic Field Laboratory, which is supported by National Science Foundation Cooperative Agreement Nos. DMR-1157490 and DMR-1644779, and the State of Florida.
\end{acknowledgments}


\begin{thebibliography}{89}
\expandafter\ifx\csname natexlab\endcsname\relax\def\natexlab#1{#1}\fi
\expandafter\ifx\csname bibnamefont\endcsname\relax
  \def\bibnamefont#1{#1}\fi
\expandafter\ifx\csname bibfnamefont\endcsname\relax
  \def\bibfnamefont#1{#1}\fi
\expandafter\ifx\csname citenamefont\endcsname\relax
  \def\citenamefont#1{#1}\fi
\expandafter\ifx\csname url\endcsname\relax
  \def\url#1{\texttt{#1}}\fi
\expandafter\ifx\csname urlprefix\endcsname\relax\def\urlprefix{URL }\fi
\providecommand{\bibinfo}[2]{#2}
\providecommand{\eprint}[2][]{\url{#2}}

\bibitem[{\citenamefont{von Klitzing et~al.}(1980)\citenamefont{von Klitzing,
  Dorda, and Pepper}}]{klitzing:1980}
\bibinfo{author}{\bibfnamefont{K.}~\bibnamefont{von Klitzing}},
  \bibinfo{author}{\bibfnamefont{G.}~\bibnamefont{Dorda}}, \bibnamefont{and}
  \bibinfo{author}{\bibfnamefont{M.}~\bibnamefont{Pepper}},
  {\bibinfo{title}{New Method for High-Accuracy Determination of the
  Fine-Structure Constant Based on Quantized Hall Resistance}},
  \bibinfo{journal}{Phys. Rev. Lett.} \textbf{\bibinfo{volume}{45}},
  \bibinfo{pages}{494} (\bibinfo{year}{1980}).

\bibitem[{\citenamefont{Tsui et~al.}(1982)\citenamefont{Tsui, Stormer, and
  Gossard}}]{tsui:1982}
\bibinfo{author}{\bibfnamefont{D.~C.} \bibnamefont{Tsui}},
  \bibinfo{author}{\bibfnamefont{H.~L.} \bibnamefont{Stormer}},
  \bibnamefont{and} \bibinfo{author}{\bibfnamefont{A.~C.}
  \bibnamefont{Gossard}}, {\bibinfo{title}{Two-Dimensional
  Magnetotransport in the Extreme Quantum Limit}}, \bibinfo{journal}{Phys. Rev.
  Lett.} \textbf{\bibinfo{volume}{48}}, \bibinfo{pages}{1559}
  (\bibinfo{year}{1982}).

\bibitem[{\citenamefont{Koulakov et~al.}(1996)\citenamefont{Koulakov, Fogler,
  and Shklovskii}}]{koulakov:1996}
\bibinfo{author}{\bibfnamefont{A.~A.} \bibnamefont{Koulakov}},
  \bibinfo{author}{\bibfnamefont{M.~M.} \bibnamefont{Fogler}},
  \bibnamefont{and} \bibinfo{author}{\bibfnamefont{B.~I.}
  \bibnamefont{Shklovskii}}, {\bibinfo{title}{Charge density wave in
  two-dimensional electron liquid in weak magnetic field}},
  \bibinfo{journal}{Phys. Rev. Lett.} \textbf{\bibinfo{volume}{76}},
  \bibinfo{pages}{499} (\bibinfo{year}{1996}).

\bibitem[{\citenamefont{Fogler et~al.}(1996)\citenamefont{Fogler, Koulakov, and
  Shklovskii}}]{fogler:1996}
\bibinfo{author}{\bibfnamefont{M.~M.} \bibnamefont{Fogler}},
  \bibinfo{author}{\bibfnamefont{A.~A.} \bibnamefont{Koulakov}},
  \bibnamefont{and} \bibinfo{author}{\bibfnamefont{B.~I.}
  \bibnamefont{Shklovskii}}, {\bibinfo{title}{Ground state of a
  two-dimensional electron liquid in a weak magnetic field}},
  \bibinfo{journal}{Phys. Rev. B} \textbf{\bibinfo{volume}{54}},
  \bibinfo{pages}{1853} (\bibinfo{year}{1996}).

\bibitem[{\citenamefont{Moessner and Chalker}(1996)}]{moessner:1996}
\bibinfo{author}{\bibfnamefont{R.}~\bibnamefont{Moessner}} \bibnamefont{and}
  \bibinfo{author}{\bibfnamefont{J.~T.} \bibnamefont{Chalker}},
  {\bibinfo{title}{Exact results for interacting electrons in high Landau
  levels}}, \bibinfo{journal}{Phys. Rev. B} \textbf{\bibinfo{volume}{54}},
  \bibinfo{pages}{5006} (\bibinfo{year}{1996}).

\bibitem[{\citenamefont{Lilly et~al.}(1999{\natexlab{a}})\citenamefont{Lilly,
  Cooper, Eisenstein, Pfeiffer, and West}}]{lilly:1999a}
\bibinfo{author}{\bibfnamefont{M.~P.} \bibnamefont{Lilly}},
  \bibinfo{author}{\bibfnamefont{K.~B.} \bibnamefont{Cooper}},
  \bibinfo{author}{\bibfnamefont{J.~P.} \bibnamefont{Eisenstein}},
  \bibinfo{author}{\bibfnamefont{L.~N.} \bibnamefont{Pfeiffer}},
  \bibnamefont{and} \bibinfo{author}{\bibfnamefont{K.~W.} \bibnamefont{West}},
  {\bibinfo{title}{Evidence for an Anisotropic State of Two-Dimensional
  Electrons in High Landau Levels}}, \bibinfo{journal}{Phys. Rev. Lett.}
  \textbf{\bibinfo{volume}{82}}, \bibinfo{pages}{394}
  (\bibinfo{year}{1999}{\natexlab{a}}).

\bibitem[{\citenamefont{Du et~al.}(1999)\citenamefont{Du, Tsui, Stormer,
  Pfeiffer, Baldwin, and West}}]{du:1999}
\bibinfo{author}{\bibfnamefont{R.~R.} \bibnamefont{Du}},
  \bibinfo{author}{\bibfnamefont{D.~C.} \bibnamefont{Tsui}},
  \bibinfo{author}{\bibfnamefont{H.~L.} \bibnamefont{Stormer}},
  \bibinfo{author}{\bibfnamefont{L.~N.} \bibnamefont{Pfeiffer}},
  \bibinfo{author}{\bibfnamefont{K.~W.} \bibnamefont{Baldwin}},
  \bibnamefont{and} \bibinfo{author}{\bibfnamefont{K.~W.} \bibnamefont{West}},
  {\bibinfo{title}{Strongly anisotropic transport in higher
  two-dimensional Landau levels}}, \bibinfo{journal}{Solid State Commun.}
  \textbf{\bibinfo{volume}{109}}, \bibinfo{pages}{389} (\bibinfo{year}{1999}).

\bibitem[{\citenamefont{Nelson et~al.}(1992)\citenamefont{Nelson, Ismail,
  Nocera, Fang, Mendez, Chu, and Meyerson}}]{nelson:1992}
\bibinfo{author}{\bibfnamefont{S.~F.} \bibnamefont{Nelson}},
  \bibinfo{author}{\bibfnamefont{K.}~\bibnamefont{Ismail}},
  \bibinfo{author}{\bibfnamefont{J.~J.} \bibnamefont{Nocera}},
  \bibinfo{author}{\bibfnamefont{F.~F.} \bibnamefont{Fang}},
  \bibinfo{author}{\bibfnamefont{E.~E.} \bibnamefont{Mendez}},
  \bibinfo{author}{\bibfnamefont{J.~O.} \bibnamefont{Chu}}, \bibnamefont{and}
  \bibinfo{author}{\bibfnamefont{B.~S.} \bibnamefont{Meyerson}},
  {\bibinfo{title}{Observation of the fractional quantum Hall effect in
  Si/SiGe heterostructures}}, \bibinfo{journal}{Appl. Phys. Lett.}
  \textbf{\bibinfo{volume}{61}}, \bibinfo{pages}{64} (\bibinfo{year}{1992}).

\bibitem[{\citenamefont{Kott et~al.}(2014)\citenamefont{Kott, Hu, Brown, and
  Kane}}]{kott:2014}
\bibinfo{author}{\bibfnamefont{T.~M.} \bibnamefont{Kott}},
  \bibinfo{author}{\bibfnamefont{B.}~\bibnamefont{Hu}},
  \bibinfo{author}{\bibfnamefont{S.~H.} \bibnamefont{Brown}}, \bibnamefont{and}
  \bibinfo{author}{\bibfnamefont{B.~E.} \bibnamefont{Kane}},
  {\bibinfo{title}{Valley-degenerate two-dimensional electrons in the
  lowest Landau level}}, \bibinfo{journal}{Phys. Rev. B}
  \textbf{\bibinfo{volume}{89}}, \bibinfo{pages}{041107(R)}
  (\bibinfo{year}{2014}).

\bibitem[{\citenamefont{De~Poortere et~al.}(2002)\citenamefont{De~Poortere,
  Shkolnikov, Tutuc, Papadakis, Shayegan, Palm, and Murphy}}]{poortere:2002}
\bibinfo{author}{\bibfnamefont{E.~P.} \bibnamefont{De~Poortere}},
  \bibinfo{author}{\bibfnamefont{Y.~P.} \bibnamefont{Shkolnikov}},
  \bibinfo{author}{\bibfnamefont{E.}~\bibnamefont{Tutuc}},
  \bibinfo{author}{\bibfnamefont{S.~J.} \bibnamefont{Papadakis}},
  \bibinfo{author}{\bibfnamefont{M.}~\bibnamefont{Shayegan}},
  \bibinfo{author}{\bibfnamefont{E.}~\bibnamefont{Palm}}, \bibnamefont{and}
  \bibinfo{author}{\bibfnamefont{T.}~\bibnamefont{Murphy}},
  {\bibinfo{title}{Enhanced electron mobility and high order fractional
  quantum Hall states in AlAs quantum wells}}, \bibinfo{journal}{Appl. Phys.
  Lett.} \textbf{\bibinfo{volume}{80}}, \bibinfo{pages}{1583}
  (\bibinfo{year}{2002}).

\bibitem[{\citenamefont{Manfra et~al.}(2002)\citenamefont{Manfra, Weimann, Hsu,
  Pfeiffer, West, Syed, Stormer, Pan, Lang, Chu et~al.}}]{manfra:2002}
\bibinfo{author}{\bibfnamefont{M.~J.} \bibnamefont{Manfra}},
  \bibinfo{author}{\bibfnamefont{N.~G.} \bibnamefont{Weimann}},
  \bibinfo{author}{\bibfnamefont{J.~W.~P.} \bibnamefont{Hsu}},
  \bibinfo{author}{\bibfnamefont{L.~N.} \bibnamefont{Pfeiffer}},
  \bibinfo{author}{\bibfnamefont{K.~W.} \bibnamefont{West}},
  \bibinfo{author}{\bibfnamefont{S.}~\bibnamefont{Syed}},
  \bibinfo{author}{\bibfnamefont{H.~L.} \bibnamefont{Stormer}},
  \bibinfo{author}{\bibfnamefont{W.}~\bibnamefont{Pan}},
  \bibinfo{author}{\bibfnamefont{D.~V.} \bibnamefont{Lang}},
  \bibinfo{author}{\bibfnamefont{S.~N.~G.} \bibnamefont{Chu}},
  \bibnamefont{et~al.}, {\bibinfo{title}{High mobility AlGaN/GaN
  heterostructures grown by plasma-assisted molecular beam epitaxy on
  semi-insulating GaN templates prepared by hydride vapor phase epitaxy}},
  \bibinfo{journal}{J. Appl. Phys.} \textbf{\bibinfo{volume}{92}},
  \bibinfo{pages}{338} (\bibinfo{year}{2002}).


\bibitem[{\citenamefont{Du et~al.}(2009)\citenamefont{Du, Skachko, Duerr,
  Luican, and Andrei}}]{du:2009}
\bibinfo{author}{\bibfnamefont{X.}~\bibnamefont{Du}},
  \bibinfo{author}{\bibfnamefont{I.}~\bibnamefont{Skachko}},
  \bibinfo{author}{\bibfnamefont{F.}~\bibnamefont{Duerr}},
  \bibinfo{author}{\bibfnamefont{A.}~\bibnamefont{Luican}}, \bibnamefont{and}
  \bibinfo{author}{\bibfnamefont{E.~Y.} \bibnamefont{Andrei}},
  {\bibinfo{title}{Fractional quantum Hall effect and insulating phase of
  Dirac electrons in graphene}}, \bibinfo{journal}{Nature}
  \textbf{\bibinfo{volume}{462}}, \bibinfo{pages}{192} (\bibinfo{year}{2009}).

\bibitem[{\citenamefont{Bolotin et~al.}(2009)\citenamefont{Bolotin, Ghahari,
  Shulman, Stormer, and Kim}}]{bolotin:2009}
\bibinfo{author}{\bibfnamefont{K.~I.} \bibnamefont{Bolotin}},
  \bibinfo{author}{\bibfnamefont{F.}~\bibnamefont{Ghahari}},
  \bibinfo{author}{\bibfnamefont{M.~D.} \bibnamefont{Shulman}},
  \bibinfo{author}{\bibfnamefont{H.~L.} \bibnamefont{Stormer}},
  \bibnamefont{and} \bibinfo{author}{\bibfnamefont{P.}~\bibnamefont{Kim}},
  {\bibinfo{title}{Observation of the fractional quantum Hall effect in
  graphene}}, \bibinfo{journal}{Nature} \textbf{\bibinfo{volume}{462}},
  \bibinfo{pages}{196} (\bibinfo{year}{2009}).

\bibitem[{\citenamefont{Piot et~al.}(2010)\citenamefont{Piot, Kunc, Potemski,
  Maude, Betthausen, Vogl, Weiss, Karczewski, and Wojtowicz}}]{piot:2010}
\bibinfo{author}{\bibfnamefont{B.~A.} \bibnamefont{Piot}},
  \bibinfo{author}{\bibfnamefont{J.}~\bibnamefont{Kunc}},
  \bibinfo{author}{\bibfnamefont{M.}~\bibnamefont{Potemski}},
  \bibinfo{author}{\bibfnamefont{D.~K.} \bibnamefont{Maude}},
  \bibinfo{author}{\bibfnamefont{C.}~\bibnamefont{Betthausen}},
  \bibinfo{author}{\bibfnamefont{A.}~\bibnamefont{Vogl}},
  \bibinfo{author}{\bibfnamefont{D.}~\bibnamefont{Weiss}},
  \bibinfo{author}{\bibfnamefont{G.}~\bibnamefont{Karczewski}},
  \bibnamefont{and}
  \bibinfo{author}{\bibfnamefont{T.}~\bibnamefont{Wojtowicz}},
  {\bibinfo{title}{Fractional quantum Hall effect in CdTe}},
  \bibinfo{journal}{Phys. Rev. B} \textbf{\bibinfo{volume}{82}},
  \bibinfo{pages}{081307(R)} (\bibinfo{year}{2010}).

\bibitem[{\citenamefont{Tsukazaki et~al.}(2010)\citenamefont{Tsukazaki,
  Akasaka, Nakahara, Ohno, Ohno, Maryenko, Ohtomo, and
  Kawasaki}}]{tsukazaki:2010}
\bibinfo{author}{\bibfnamefont{A.}~\bibnamefont{Tsukazaki}},
  \bibinfo{author}{\bibfnamefont{S.}~\bibnamefont{Akasaka}},
  \bibinfo{author}{\bibfnamefont{K.}~\bibnamefont{Nakahara}},
  \bibinfo{author}{\bibfnamefont{Y.}~\bibnamefont{Ohno}},
  \bibinfo{author}{\bibfnamefont{H.}~\bibnamefont{Ohno}},
  \bibinfo{author}{\bibfnamefont{D.}~\bibnamefont{Maryenko}},
  \bibinfo{author}{\bibfnamefont{A.}~\bibnamefont{Ohtomo}}, \bibnamefont{and}
  \bibinfo{author}{\bibfnamefont{M.}~\bibnamefont{Kawasaki}},
  {\bibinfo{title}{Observation of the fractional quantum Hall effect in an
  oxide}}, \bibinfo{journal}{Nature Mat.} \textbf{\bibinfo{volume}{9}},
  \bibinfo{pages}{889} (\bibinfo{year}{2010}).

\bibitem[{\citenamefont{Shi et~al.}(2015)\citenamefont{Shi, Zudov, Morrison,
  and Myronov}}]{shi:2015c}
\bibinfo{author}{\bibfnamefont{Q.}~\bibnamefont{Shi}},
  \bibinfo{author}{\bibfnamefont{M.~A.} \bibnamefont{Zudov}},
  \bibinfo{author}{\bibfnamefont{C.}~\bibnamefont{Morrison}}, \bibnamefont{and}
  \bibinfo{author}{\bibfnamefont{M.}~\bibnamefont{Myronov}},
  {\bibinfo{title}{Spinless composite fermions in an ultrahigh-quality
  strained Ge quantum well}}, \bibinfo{journal}{Phys. Rev. B}
  \textbf{\bibinfo{volume}{91}}, \bibinfo{pages}{241303(R)}
  (\bibinfo{year}{2015}).

\bibitem[{\citenamefont{Ma et~al.}(2017)\citenamefont{Ma, Hossain,
  Villegas~Rosales, Deng, Tschirky, Wegscheider, and Shayegan}}]{ma:2017}
\bibinfo{author}{\bibfnamefont{M.~K.} \bibnamefont{Ma}},
  \bibinfo{author}{\bibfnamefont{M.~S.} \bibnamefont{Hossain}},
  \bibinfo{author}{\bibfnamefont{K.~A.} \bibnamefont{Villegas~Rosales}},
  \bibinfo{author}{\bibfnamefont{H.}~\bibnamefont{Deng}},
  \bibinfo{author}{\bibfnamefont{T.}~\bibnamefont{Tschirky}},
  \bibinfo{author}{\bibfnamefont{W.}~\bibnamefont{Wegscheider}},
  \bibnamefont{and} \bibinfo{author}{\bibfnamefont{M.}~\bibnamefont{Shayegan}},
  {\bibinfo{title}{Observation of fractional quantum Hall effect in an
  InAs quantum well}}, \bibinfo{journal}{Phys. Rev. B}
  \textbf{\bibinfo{volume}{96}}, \bibinfo{pages}{241301(R)}
  (\bibinfo{year}{2017}).

\bibitem[{not({\natexlab{a}})}]{note:nem}
\bibinfo{note}{Apart from QHSs in GaAs, electron nematic states were identified
  in Sr$_3$Ru$_2$O$_7$ \citep{borzi:2007}, high-T$_{\rm C}$ superconductors
  \citep{daou:2010,chu:2010}, URu$_2$Si$_2$ \citep{okazaki:2011},
  WSe$_2$/WS$_2$ \citep{jin:2020}, and twisted bilayer graphene
  \citep{cao:2020}.}

\bibitem[{not({\natexlab{b}})}]{note:fradkin}
\bibinfo{note}{Considering thermal and quantum fluctuations, several electron
  liquid crystal-like phases have also been proposed \citep{fradkin:1999}.}

\bibitem[{\citenamefont{Fil}(2000)}]{fil:2000}
\bibinfo{author}{\bibfnamefont{D.~V.} \bibnamefont{Fil}},
  {\bibinfo{title}{Piezoelectric mechanism for the orientation of stripe
  structures in two-dimensional electron systems}}, \bibinfo{journal}{Low Temp.
  Phys.} \textbf{\bibinfo{volume}{26}}, \bibinfo{pages}{581}
  (\bibinfo{year}{2000}).

\bibitem[{\citenamefont{Koduvayur et~al.}(2011)\citenamefont{Koduvayur,
  Lyanda-Geller, Khlebnikov, Cs\'athy, Manfra, Pfeiffer, West, and
  Rokhinson}}]{koduvayur:2011}
\bibinfo{author}{\bibfnamefont{S.~P.} \bibnamefont{Koduvayur}},
  \bibinfo{author}{\bibfnamefont{Y.}~\bibnamefont{Lyanda-Geller}},
  \bibinfo{author}{\bibfnamefont{S.}~\bibnamefont{Khlebnikov}},
  \bibinfo{author}{\bibfnamefont{G.}~\bibnamefont{Cs\'athy}},
  \bibinfo{author}{\bibfnamefont{M.~J.} \bibnamefont{Manfra}},
  \bibinfo{author}{\bibfnamefont{L.~N.} \bibnamefont{Pfeiffer}},
  \bibinfo{author}{\bibfnamefont{K.~W.} \bibnamefont{West}}, \bibnamefont{and}
  \bibinfo{author}{\bibfnamefont{L.~P.} \bibnamefont{Rokhinson}},
  {\bibinfo{title}{Effect of Strain on Stripe Phases in the Quantum Hall
  Regime}}, \bibinfo{journal}{Phys. Rev. Lett.} \textbf{\bibinfo{volume}{106}},
  \bibinfo{pages}{016804} (\bibinfo{year}{2011}).

\bibitem[{\citenamefont{{Sodemann} and {MacDonald}}(2013)}]{sodemann:2013}
\bibinfo{author}{\bibfnamefont{I.}~\bibnamefont{{Sodemann}}} \bibnamefont{and}
  \bibinfo{author}{\bibfnamefont{A.~H.} \bibnamefont{{MacDonald}}},
  {\bibinfo{title}{Theory of Native Orientational Pinning in Quantum Hall
  Nematics}}, \bibinfo{journal}{arXiv:1307.5489}
  (\bibinfo{year}{2013}).

\bibitem[{\citenamefont{Pollanen et~al.}(2015)\citenamefont{Pollanen, Cooper,
  Brandsen, Eisenstein, Pfeiffer, and West}}]{pollanen:2015}
\bibinfo{author}{\bibfnamefont{J.}~\bibnamefont{Pollanen}},
  \bibinfo{author}{\bibfnamefont{K.~B.} \bibnamefont{Cooper}},
  \bibinfo{author}{\bibfnamefont{S.}~\bibnamefont{Brandsen}},
  \bibinfo{author}{\bibfnamefont{J.~P.} \bibnamefont{Eisenstein}},
  \bibinfo{author}{\bibfnamefont{L.~N.} \bibnamefont{Pfeiffer}},
  \bibnamefont{and} \bibinfo{author}{\bibfnamefont{K.~W.} \bibnamefont{West}},
  {\bibinfo{title}{Heterostructure symmetry and the orientation of the
  quantum Hall nematic phases}}, \bibinfo{journal}{Phys. Rev. B}
  \textbf{\bibinfo{volume}{92}}, \bibinfo{pages}{115410}
  (\bibinfo{year}{2015}).

\bibitem[{not({\natexlab{c}})}]{note:ex}
\bibinfo{note}{QHSs were found to align along the $\hard$ direction in a tunable
  density heterostructure insulated gate field effect transistor at densities
  $\ne$ between $3\times 10^{11}$ cm$^{-2}$ and $4.6\times 10^{11}$ cm$^{-2}$
  \citep{zhu:2002} and in single heterointerface devices with $\ne \simeq 2
  \times 10^{11}$ cm$^{-2}$ and cap thickness exceeding 1 $\mu$m
  \citep{pollanen:2015}.}

\bibitem[{not({\natexlab{d}})}]{note:ts}
\bibinfo{note}{Indeed, anisotropies emerging under the in-plane magnetic field,
  which is known to provide symmetry breaking
  \citep{pan:1999,lilly:1999b,jungwirth:1999,stanescu:2000,zhu:2009,shi:2016c,shi:2017c},
  have been observed in ZnO \citep{falson:2018} and AlAs \citep{hossain:2018}.}

\bibitem[{\citenamefont{Cooper et~al.}(1999)\citenamefont{Cooper, Lilly,
  Eisenstein, Pfeiffer, and West}}]{cooper:1999}
\bibinfo{author}{\bibfnamefont{K.~B.} \bibnamefont{Cooper}},
  \bibinfo{author}{\bibfnamefont{M.~P.} \bibnamefont{Lilly}},
  \bibinfo{author}{\bibfnamefont{J.~P.} \bibnamefont{Eisenstein}},
  \bibinfo{author}{\bibfnamefont{L.~N.} \bibnamefont{Pfeiffer}},
  \bibnamefont{and} \bibinfo{author}{\bibfnamefont{K.~W.} \bibnamefont{West}},
  {\bibinfo{title}{Insulating phases of two-dimensional electrons in high
  Landau levels: Observation of sharp thresholds to conduction}},
  \bibinfo{journal}{Phys. Rev. B} \textbf{\bibinfo{volume}{60}},
  \bibinfo{pages}{R11285} (\bibinfo{year}{1999}).

\bibitem[{\citenamefont{Eisenstein et~al.}(2002)\citenamefont{Eisenstein,
  Cooper, Pfeiffer, and West}}]{eisenstein:2002}
\bibinfo{author}{\bibfnamefont{J.~P.} \bibnamefont{Eisenstein}},
  \bibinfo{author}{\bibfnamefont{K.~B.} \bibnamefont{Cooper}},
  \bibinfo{author}{\bibfnamefont{L.~N.} \bibnamefont{Pfeiffer}},
  \bibnamefont{and} \bibinfo{author}{\bibfnamefont{K.~W.} \bibnamefont{West}},
  {\bibinfo{title}{Insulating and Fractional Quantum Hall States in the
  First Excited Landau Level}}, \bibinfo{journal}{Phys. Rev. Lett.}
  \textbf{\bibinfo{volume}{88}}, \bibinfo{pages}{076801}
  (\bibinfo{year}{2002}).

\bibitem[{\citenamefont{Lewis et~al.}(2002)\citenamefont{Lewis, Ye, Engel,
  Tsui, Pfeiffer, and West}}]{lewis:2002}
\bibinfo{author}{\bibfnamefont{R.~M.} \bibnamefont{Lewis}},
  \bibinfo{author}{\bibfnamefont{P.~D.} \bibnamefont{Ye}},
  \bibinfo{author}{\bibfnamefont{L.~W.} \bibnamefont{Engel}},
  \bibinfo{author}{\bibfnamefont{D.~C.} \bibnamefont{Tsui}},
  \bibinfo{author}{\bibfnamefont{L.~N.} \bibnamefont{Pfeiffer}},
  \bibnamefont{and} \bibinfo{author}{\bibfnamefont{K.~W.} \bibnamefont{West}},
  {\bibinfo{title}{Microwave Resonance of the Bubble Phases in 1/4 and 3/4
  Filled High Landau Levels}}, \bibinfo{journal}{Phys. Rev. Lett.}
  \textbf{\bibinfo{volume}{89}}, \bibinfo{pages}{136804}
  (\bibinfo{year}{2002}).

\bibitem[{\citenamefont{Deng et~al.}(2012)\citenamefont{Deng, Kumar, Manfra,
  Pfeiffer, West, and Cs\'athy}}]{deng:2012a}
\bibinfo{author}{\bibfnamefont{N.}~\bibnamefont{Deng}},
  \bibinfo{author}{\bibfnamefont{A.}~\bibnamefont{Kumar}},
  \bibinfo{author}{\bibfnamefont{M.~J.} \bibnamefont{Manfra}},
  \bibinfo{author}{\bibfnamefont{L.~N.} \bibnamefont{Pfeiffer}},
  \bibinfo{author}{\bibfnamefont{K.~W.} \bibnamefont{West}}, \bibnamefont{and}
  \bibinfo{author}{\bibfnamefont{G.~A.} \bibnamefont{Cs\'athy}},
  {\bibinfo{title}{Collective Nature of the Reentrant Integer Quantum Hall
  States in the Second Landau Level}}, \bibinfo{journal}{Phys. Rev. Lett.}
  \textbf{\bibinfo{volume}{108}}, \bibinfo{pages}{086803}
  (\bibinfo{year}{2012}).

\bibitem[{\citenamefont{Rossokhaty et~al.}(2016)\citenamefont{Rossokhaty, Baum,
  Folk, Watson, Gardner, and Manfra}}]{rossokhaty:2016}
\bibinfo{author}{\bibfnamefont{A.~V.} \bibnamefont{Rossokhaty}},
  \bibinfo{author}{\bibfnamefont{Y.}~\bibnamefont{Baum}},
  \bibinfo{author}{\bibfnamefont{J.~A.} \bibnamefont{Folk}},
  \bibinfo{author}{\bibfnamefont{J.~D.} \bibnamefont{Watson}},
  \bibinfo{author}{\bibfnamefont{G.~C.} \bibnamefont{Gardner}},
  \bibnamefont{and} \bibinfo{author}{\bibfnamefont{M.~J.}
  \bibnamefont{Manfra}}, {\bibinfo{title}{Electron-Hole Asymmetric Chiral
  Breakdown of Reentrant Quantum Hall States}}, \bibinfo{journal}{Phys. Rev.
  Lett.} \textbf{\bibinfo{volume}{117}}, \bibinfo{pages}{166805}
  (\bibinfo{year}{2016}).

\bibitem[{\citenamefont{Friess et~al.}(2017)\citenamefont{Friess, Peng,
  Rosenow, von Oppen, Umansky, von Klitzing, and Smet}}]{friess:2017}
\bibinfo{author}{\bibfnamefont{B.}~\bibnamefont{Friess}},
  \bibinfo{author}{\bibfnamefont{Y.}~\bibnamefont{Peng}},
  \bibinfo{author}{\bibfnamefont{B.}~\bibnamefont{Rosenow}},
  \bibinfo{author}{\bibfnamefont{F.}~\bibnamefont{von Oppen}},
  \bibinfo{author}{\bibfnamefont{V.}~\bibnamefont{Umansky}},
  \bibinfo{author}{\bibfnamefont{K.}~\bibnamefont{von Klitzing}},
  \bibnamefont{and} \bibinfo{author}{\bibfnamefont{J.~H.} \bibnamefont{Smet}},
  {\bibinfo{title}{Negative permittivity in bubble and stripe phases}},
  \bibinfo{journal}{Nature Phys.} \textbf{\bibinfo{volume}{13}},
  \bibinfo{pages}{1124} (\bibinfo{year}{2017}).

\bibitem[{\citenamefont{Bennaceur et~al.}(2018)\citenamefont{Bennaceur, Lupien,
  Reulet, Gervais, Pfeiffer, and West}}]{bennaceur:2018}
\bibinfo{author}{\bibfnamefont{K.}~\bibnamefont{Bennaceur}},
  \bibinfo{author}{\bibfnamefont{C.}~\bibnamefont{Lupien}},
  \bibinfo{author}{\bibfnamefont{B.}~\bibnamefont{Reulet}},
  \bibinfo{author}{\bibfnamefont{G.}~\bibnamefont{Gervais}},
  \bibinfo{author}{\bibfnamefont{L.~N.} \bibnamefont{Pfeiffer}},
  \bibnamefont{and} \bibinfo{author}{\bibfnamefont{K.~W.} \bibnamefont{West}},
  {\bibinfo{title}{Competing Charge Density Waves Probed by Nonlinear
  Transport and Noise in the Second and Third Landau Levels}},
  \bibinfo{journal}{Phys. Rev. Lett.} \textbf{\bibinfo{volume}{120}},
  \bibinfo{pages}{136801} (\bibinfo{year}{2018}).

\bibitem[{\citenamefont{Friess et~al.}(2018)\citenamefont{Friess, Umansky, von
  Klitzing, and Smet}}]{friess:2018}
\bibinfo{author}{\bibfnamefont{B.}~\bibnamefont{Friess}},
  \bibinfo{author}{\bibfnamefont{V.}~\bibnamefont{Umansky}},
  \bibinfo{author}{\bibfnamefont{K.}~\bibnamefont{von Klitzing}},
  \bibnamefont{and} \bibinfo{author}{\bibfnamefont{J.~H.} \bibnamefont{Smet}},
  {\bibinfo{title}{Current Flow in the Bubble and Stripe Phases}},
  \bibinfo{journal}{Phys. Rev. Lett.} \textbf{\bibinfo{volume}{120}},
  \bibinfo{pages}{137603} (\bibinfo{year}{2018}).

\bibitem[{\citenamefont{Fu et~al.}(2019)\citenamefont{Fu, Shi, Zudov, Gardner,
  Watson, and Manfra}}]{fu:2019}
\bibinfo{author}{\bibfnamefont{X.}~\bibnamefont{Fu}},
  \bibinfo{author}{\bibfnamefont{Q.}~\bibnamefont{Shi}},
  \bibinfo{author}{\bibfnamefont{M.~A.} \bibnamefont{Zudov}},
  \bibinfo{author}{\bibfnamefont{G.~C.} \bibnamefont{Gardner}},
  \bibinfo{author}{\bibfnamefont{J.~D.} \bibnamefont{Watson}},
  \bibnamefont{and} \bibinfo{author}{\bibfnamefont{M.~J.}
  \bibnamefont{Manfra}}, {\bibinfo{title}{Two- and three-electron bubbles
  in
  ${\mathrm{Al}}_{x}{\mathrm{Ga}}_{1\ensuremath{-}x}\mathrm{As}$/${\mathrm{Al}}_{0.24}{\mathrm{Ga}}_{0.76}\mathrm{As}$
  quantum wells}}, \bibinfo{journal}{Phys. Rev. B}
  \textbf{\bibinfo{volume}{99}}, \bibinfo{pages}{161402(R)}
  (\bibinfo{year}{2019}).

\bibitem[{\citenamefont{Ro et~al.}(2019)\citenamefont{Ro, Deng, Watson, Manfra,
  Pfeiffer, West, and Cs{\'a}thy}}]{ro:2019}
\bibinfo{author}{\bibfnamefont{D.}~\bibnamefont{Ro}},
  \bibinfo{author}{\bibfnamefont{N.}~\bibnamefont{Deng}},
  \bibinfo{author}{\bibfnamefont{J.~D.} \bibnamefont{Watson}},
  \bibinfo{author}{\bibfnamefont{M.~J.} \bibnamefont{Manfra}},
  \bibinfo{author}{\bibfnamefont{L.~N.} \bibnamefont{Pfeiffer}},
  \bibinfo{author}{\bibfnamefont{K.~W.} \bibnamefont{West}}, \bibnamefont{and}
  \bibinfo{author}{\bibfnamefont{G.~A.} \bibnamefont{Cs{\'a}thy}},
  {\bibinfo{title}{Electron bubbles and the structure of the orbital wave
  function}}, \bibinfo{journal}{Phys. Rev. B} \textbf{\bibinfo{volume}{99}},
  \bibinfo{pages}{201111(R)} (\bibinfo{year}{2019}).

\bibitem[{\citenamefont{Chen et~al.}(2019)\citenamefont{Chen, Ribeiro-Palau,
  Yang, Watanabe, Taniguchi, Hone, Goerbig, and Dean}}]{chen:2019}
\bibinfo{author}{\bibfnamefont{S.}~\bibnamefont{Chen}},
  \bibinfo{author}{\bibfnamefont{R.}~\bibnamefont{Ribeiro-Palau}},
  \bibinfo{author}{\bibfnamefont{K.}~\bibnamefont{Yang}},
  \bibinfo{author}{\bibfnamefont{K.}~\bibnamefont{Watanabe}},
  \bibinfo{author}{\bibfnamefont{T.}~\bibnamefont{Taniguchi}},
  \bibinfo{author}{\bibfnamefont{J.}~\bibnamefont{Hone}},
  \bibinfo{author}{\bibfnamefont{M.~O.} \bibnamefont{Goerbig}},
  \bibnamefont{and} \bibinfo{author}{\bibfnamefont{C.~R.} \bibnamefont{Dean}},
  {\bibinfo{title}{Competing Fractional Quantum Hall and Electron Solid
  Phases in Graphene}}, \bibinfo{journal}{Phys. Rev. Lett.}
  \textbf{\bibinfo{volume}{122}}, \bibinfo{pages}{026802}
  (\bibinfo{year}{2019}).

\bibitem[{\citenamefont{Huang et~al.}(2020)\citenamefont{Huang, Sammon, Zudov,
  and Shklovskii}}]{huang:2020}
\bibinfo{author}{\bibfnamefont{Y.}~\bibnamefont{Huang}},
  \bibinfo{author}{\bibfnamefont{M.}~\bibnamefont{Sammon}},
  \bibinfo{author}{\bibfnamefont{M.~A.} \bibnamefont{Zudov}}, \bibnamefont{and}
  \bibinfo{author}{\bibfnamefont{B.~I.} \bibnamefont{Shklovskii}},
  {\bibinfo{title}{Isotropically conducting (hidden) quantum Hall stripe
  phases in a two-dimensional electron gas}}, \bibinfo{journal}{Phys. Rev. B}
  \textbf{\bibinfo{volume}{101}}, \bibinfo{pages}{161302(R)}
  (\bibinfo{year}{2020}).

\bibitem[{\citenamefont{MacDonald and Fisher}(2000)}]{macdonald:2000}
\bibinfo{author}{\bibfnamefont{A.~H.} \bibnamefont{MacDonald}}
  \bibnamefont{and} \bibinfo{author}{\bibfnamefont{M.~P.~A.}
  \bibnamefont{Fisher}}, {\bibinfo{title}{Quantum theory of quantum Hall
  smectics}}, \bibinfo{journal}{Phys. Rev. B} \textbf{\bibinfo{volume}{61}},
  \bibinfo{pages}{5724} (\bibinfo{year}{2000}).

\bibitem[{\citenamefont{von Oppen et~al.}(2000)\citenamefont{von Oppen,
  Halperin, and Stern}}]{oppen:2000}
\bibinfo{author}{\bibfnamefont{F.}~\bibnamefont{von Oppen}},
  \bibinfo{author}{\bibfnamefont{B.~I.} \bibnamefont{Halperin}},
  \bibnamefont{and} \bibinfo{author}{\bibfnamefont{A.}~\bibnamefont{Stern}},
  {\bibinfo{title}{Conductivity Tensor of Striped Quantum Hall Phases}},
  \bibinfo{journal}{Phys. Rev. Lett.} \textbf{\bibinfo{volume}{84}},
  \bibinfo{pages}{2937} (\bibinfo{year}{2000}).

\bibitem[{\citenamefont{Sammon et~al.}(2019)\citenamefont{Sammon, Fu, Huang,
  Zudov, Shklovskii, Gardner, Watson, Manfra, Baldwin, Pfeiffer
  et~al.}}]{sammon:2019}
\bibinfo{author}{\bibfnamefont{M.}~\bibnamefont{Sammon}},
  \bibinfo{author}{\bibfnamefont{X.}~\bibnamefont{Fu}},
  \bibinfo{author}{\bibfnamefont{Y.}~\bibnamefont{Huang}},
  \bibinfo{author}{\bibfnamefont{M.~A.} \bibnamefont{Zudov}},
  \bibinfo{author}{\bibfnamefont{B.~I.} \bibnamefont{Shklovskii}},
  \bibinfo{author}{\bibfnamefont{G.~C.} \bibnamefont{Gardner}},
  \bibinfo{author}{\bibfnamefont{J.~D.} \bibnamefont{Watson}},
  \bibinfo{author}{\bibfnamefont{M.~J.} \bibnamefont{Manfra}},
  \bibinfo{author}{\bibfnamefont{K.~W.} \bibnamefont{Baldwin}},
  \bibinfo{author}{\bibfnamefont{L.~N.} \bibnamefont{Pfeiffer}},
  \bibnamefont{et~al.}, {\bibinfo{title}{Resistivity anisotropy of quantum
  Hall stripe phases}}, \bibinfo{journal}{Phys. Rev. B}
  \textbf{\bibinfo{volume}{100}}, \bibinfo{pages}{241303(R)}
  (\bibinfo{year}{2019}).

\bibitem[{not({\natexlab{e}})}]{note:ly}
\bibinfo{note}{The mean free path along stripes limited by isotropic impurity
  scattering can be calculated using $L_y^2 = (1/2\pi) \int_{0}^{2\pi} d\theta
  (2R_c \cos \theta)^2 = 2 R_c^2$.}

\bibitem[{\citenamefont{Ando and Uemura}(1974)}]{ando:1974.a}
\bibinfo{author}{\bibfnamefont{T.}~\bibnamefont{Ando}} \bibnamefont{and}
  \bibinfo{author}{\bibfnamefont{Y.}~\bibnamefont{Uemura}},
  {\bibinfo{title}{Theory of Quantum Transport in a Two-Dimensional
  Electron System under Magnetic Fields. I. Characteristics of Level Broadening
  and Transport under Strong Fields}}, \bibinfo{journal}{J. Phys. Soc. Jpn.}
  \textbf{\bibinfo{volume}{36}}, \bibinfo{pages}{959} (\bibinfo{year}{1974}).

\bibitem[{\citenamefont{Coleridge et~al.}(1994)\citenamefont{Coleridge,
  Zawadzki, and Sachrajda}}]{coleridge:1994}
\bibinfo{author}{\bibfnamefont{P.~T.} \bibnamefont{Coleridge}},
  \bibinfo{author}{\bibfnamefont{P.}~\bibnamefont{Zawadzki}}, \bibnamefont{and}
  \bibinfo{author}{\bibfnamefont{A.~S.} \bibnamefont{Sachrajda}},
  {\bibinfo{title}{Peak values of resistivity in high-mobility
  quantum-Hall-effect samples}}, \bibinfo{journal}{Phys. Rev. B}
  \textbf{\bibinfo{volume}{49}}, \bibinfo{pages}{10798} (\bibinfo{year}{1994}).

\bibitem[{not({\natexlab{f}})}]{note:al}
\bibinfo{note}{Our estimates show that $\Delta \nu$ reaches a maximum of 6 at
  $x \approx 0.1$\,\%.}

\bibitem[{\citenamefont{Gardner et~al.}(2013)\citenamefont{Gardner, Watson,
  Mondal, Deng, Cs\'athy, and Manfra}}]{gardner:2013}
\bibinfo{author}{\bibfnamefont{G.~C.} \bibnamefont{Gardner}},
  \bibinfo{author}{\bibfnamefont{J.~D.} \bibnamefont{Watson}},
  \bibinfo{author}{\bibfnamefont{S.}~\bibnamefont{Mondal}},
  \bibinfo{author}{\bibfnamefont{N.}~\bibnamefont{Deng}},
  \bibinfo{author}{\bibfnamefont{G.~A.} \bibnamefont{Cs\'athy}},
  \bibnamefont{and} \bibinfo{author}{\bibfnamefont{M.~J.}
  \bibnamefont{Manfra}}, {\bibinfo{title}{Growth and electrical
  characterization of Al$_{0.24}$Ga$_{0.76}$As/Al$_{x}$Ga$_{1-x}$As/
  Al$_{0.24}$Ga$_{0.76}$As modulation-doped quantum wells with extremely low
  x}}, \bibinfo{journal}{Appl. Phys. Lett.} \textbf{\bibinfo{volume}{102}},
  \bibinfo{pages}{252103} (\bibinfo{year}{2013}).

\bibitem[{not({\natexlab{g}})}]{note:gamma}
\bibinfo{note}{In general, $\alpha$ should be modified by a factor
  $\sqrt{2\gamma}$, where $\gamma$ is a parameter depending on the nature of
  scattering \citep{sammon:2019}. In samples without alloy disorder, $\gamma$
  decreases with $N_1/N_2$ where $N_1$ ($N_2$) is the concentration of Coulomb
  impurities in the barrier (quantum well), starting from $\gamma \approx 0.43$
  at $N_1 = N_2$ \citep{sammon:2019}, whereas in samples where scattering is
  dominated by alloy disorder $\gamma \approx 0.53$ \citep{huang:2020}. While
  we estimate $\gamma \approx 0.4$ in our samples A and B, we will assume
  $\gamma = 0.5$ for simplicity.}

\bibitem[{\citenamefont{Shi}(2017)}]{shi:thesis}
\bibinfo{author}{\bibfnamefont{Q.}~\bibnamefont{Shi}},
  {\bibinfo{title}{Magnetotransport in quantum Hall systems at high Landau
  levels}}, Ph.D. thesis, \bibinfo{school}{University of Minnesota}, 
  \bibinfo{year}{2017}.

\bibitem[{\citenamefont{Raikh and Shahbazyan}(1993)}]{raikh:1993}
\bibinfo{author}{\bibfnamefont{M.~E.} \bibnamefont{Raikh}} \bibnamefont{and}
  \bibinfo{author}{\bibfnamefont{T.~V.} \bibnamefont{Shahbazyan}},
  {\bibinfo{title}{High Landau levels in a smooth random potential for
  two-dimensional electrons}}, \bibinfo{journal}{Phys. Rev. B}
  \textbf{\bibinfo{volume}{47}}, \bibinfo{pages}{1522} (\bibinfo{year}{1993}).

\bibitem[{\citenamefont{Mirlin et~al.}(1996)\citenamefont{Mirlin, Altshuler,
  and W\"olfle}}]{mirlin:1996}
\bibinfo{author}{\bibfnamefont{A.~D.} \bibnamefont{Mirlin}},
  \bibinfo{author}{\bibfnamefont{E.}~\bibnamefont{Altshuler}},
  \bibnamefont{and} \bibinfo{author}{\bibfnamefont{P.}~\bibnamefont{W\"olfle}},
  {\bibinfo{title}{Quasiclassical approach to impurity effect on
  magnetooscillations in 2D metals}}, \bibinfo{journal}{Ann. Phys. (N.Y.)}
  \textbf{\bibinfo{volume}{508}}, \bibinfo{pages}{281} (\bibinfo{year}{1996}).

\bibitem[{\citenamefont{Shi et~al.}(2016{\natexlab{a}})\citenamefont{Shi,
  Studenikin, Zudov, Baldwin, Pfeiffer, and West}}]{shi:2016a}
\bibinfo{author}{\bibfnamefont{Q.}~\bibnamefont{Shi}},
  \bibinfo{author}{\bibfnamefont{S.~A.} \bibnamefont{Studenikin}},
  \bibinfo{author}{\bibfnamefont{M.~A.} \bibnamefont{Zudov}},
  \bibinfo{author}{\bibfnamefont{K.~W.} \bibnamefont{Baldwin}},
  \bibinfo{author}{\bibfnamefont{L.~N.} \bibnamefont{Pfeiffer}},
  \bibnamefont{and} \bibinfo{author}{\bibfnamefont{K.~W.} \bibnamefont{West}},
  {\bibinfo{title}{Microwave photoresistance in an ultra-high-quality GaAs
  quantum well}}, \bibinfo{journal}{Phys. Rev. B}
  \textbf{\bibinfo{volume}{93}}, \bibinfo{pages}{121305(R)}
  (\bibinfo{year}{2016}{\natexlab{a}}).

\bibitem[{\citenamefont{Shi et~al.}(2017{\natexlab{a}})\citenamefont{Shi,
  Zudov, Dmitriev, Baldwin, Pfeiffer, and West}}]{shi:2017a}
\bibinfo{author}{\bibfnamefont{Q.}~\bibnamefont{Shi}},
  \bibinfo{author}{\bibfnamefont{M.~A.} \bibnamefont{Zudov}},
  \bibinfo{author}{\bibfnamefont{I.~A.} \bibnamefont{Dmitriev}},
  \bibinfo{author}{\bibfnamefont{K.~W.} \bibnamefont{Baldwin}},
  \bibinfo{author}{\bibfnamefont{L.~N.} \bibnamefont{Pfeiffer}},
  \bibnamefont{and} \bibinfo{author}{\bibfnamefont{K.~W.} \bibnamefont{West}},
  {\bibinfo{title}{Fine structure of high-power microwave-induced
  resistance oscillations}}, \bibinfo{journal}{Phys. Rev. B}
  \textbf{\bibinfo{volume}{95}}, \bibinfo{pages}{041403(R)}
  (\bibinfo{year}{2017}{\natexlab{a}}).

\bibitem[{\citenamefont{Zudov et~al.}(2017)\citenamefont{Zudov, Dmitriev,
  Friess, Shi, Umansky, von Klitzing, and Smet}}]{zudov:2017}
\bibinfo{author}{\bibfnamefont{M.~A.} \bibnamefont{Zudov}},
  \bibinfo{author}{\bibfnamefont{I.~A.} \bibnamefont{Dmitriev}},
  \bibinfo{author}{\bibfnamefont{B.}~\bibnamefont{Friess}},
  \bibinfo{author}{\bibfnamefont{Q.}~\bibnamefont{Shi}},
  \bibinfo{author}{\bibfnamefont{V.}~\bibnamefont{Umansky}},
  \bibinfo{author}{\bibfnamefont{K.}~\bibnamefont{von Klitzing}},
  \bibnamefont{and} \bibinfo{author}{\bibfnamefont{J.}~\bibnamefont{Smet}},
  {\bibinfo{title}{Hall field-induced resistance oscillations in a
  tunable-density GaAs quantum well}}, \bibinfo{journal}{Phys. Rev. B}
  \textbf{\bibinfo{volume}{96}}, \bibinfo{pages}{121301(R)}
  (\bibinfo{year}{2017}).

\bibitem[{\citenamefont{Zudov et~al.}(2001)\citenamefont{Zudov, Du, Simmons,
  and Reno}}]{zudov:2001a}
\bibinfo{author}{\bibfnamefont{M.~A.} \bibnamefont{Zudov}},
  \bibinfo{author}{\bibfnamefont{R.~R.} \bibnamefont{Du}},
  \bibinfo{author}{\bibfnamefont{J.~A.} \bibnamefont{Simmons}},
  \bibnamefont{and} \bibinfo{author}{\bibfnamefont{J.~L.} \bibnamefont{Reno}},
  {\bibinfo{title}{Shubnikov--de Haas-like oscillations in millimeterwave
  photoconductivity in a high-mobility two-dimensional electron gas}},
  \bibinfo{journal}{Phys. Rev. B} \textbf{\bibinfo{volume}{64}},
  \bibinfo{pages}{201311(R)} (\bibinfo{year}{2001}).

\bibitem[{\citenamefont{Ye et~al.}(2001)\citenamefont{Ye, Engel, Tsui, Simmons,
  Wendt, Vawter, and Reno}}]{ye:2001}
\bibinfo{author}{\bibfnamefont{P.~D.} \bibnamefont{Ye}},
  \bibinfo{author}{\bibfnamefont{L.~W.} \bibnamefont{Engel}},
  \bibinfo{author}{\bibfnamefont{D.~C.} \bibnamefont{Tsui}},
  \bibinfo{author}{\bibfnamefont{J.~A.} \bibnamefont{Simmons}},
  \bibinfo{author}{\bibfnamefont{J.~R.} \bibnamefont{Wendt}},
  \bibinfo{author}{\bibfnamefont{G.~A.} \bibnamefont{Vawter}},
  \bibnamefont{and} \bibinfo{author}{\bibfnamefont{J.~L.} \bibnamefont{Reno}},
  {\bibinfo{title}{Giant microwave photoresistance of two-dimensional
  electron gas}}, \bibinfo{journal}{Appl. Phys. Lett.}
  \textbf{\bibinfo{volume}{79}}, \bibinfo{pages}{2193} (\bibinfo{year}{2001}).

\bibitem[{\citenamefont{Dmitriev et~al.}(2012)\citenamefont{Dmitriev, Mirlin,
  Polyakov, and Zudov}}]{dmitriev:2012}
\bibinfo{author}{\bibfnamefont{I.~A.} \bibnamefont{Dmitriev}},
  \bibinfo{author}{\bibfnamefont{A.~D.} \bibnamefont{Mirlin}},
  \bibinfo{author}{\bibfnamefont{D.~G.} \bibnamefont{Polyakov}},
  \bibnamefont{and} \bibinfo{author}{\bibfnamefont{M.~A.} \bibnamefont{Zudov}},
  {\bibinfo{title}{Nonequilibrium phenomena in high Landau levels}},
  \bibinfo{journal}{Rev. Mod. Phys.} \textbf{\bibinfo{volume}{84}},
  \bibinfo{pages}{1709} (\bibinfo{year}{2012}).

\bibitem[{\citenamefont{Yang et~al.}(2002)\citenamefont{Yang, Zhang, Du,
  Simmons, and Reno}}]{yang:2002}
\bibinfo{author}{\bibfnamefont{C.~L.} \bibnamefont{Yang}},
  \bibinfo{author}{\bibfnamefont{J.}~\bibnamefont{Zhang}},
  \bibinfo{author}{\bibfnamefont{R.~R.} \bibnamefont{Du}},
  \bibinfo{author}{\bibfnamefont{J.~A.} \bibnamefont{Simmons}},
  \bibnamefont{and} \bibinfo{author}{\bibfnamefont{J.~L.} \bibnamefont{Reno}},
  {\bibinfo{title}{Zener Tunneling Between Landau Orbits in a
  High-Mobility Two-Dimensional Electron Gas}}, \bibinfo{journal}{Phys. Rev.
  Lett.} \textbf{\bibinfo{volume}{89}}, \bibinfo{pages}{076801}
  (\bibinfo{year}{2002}).

\bibitem[{\citenamefont{Zhang et~al.}(2007)\citenamefont{Zhang, Chiang, Zudov,
  Pfeiffer, and West}}]{zhang:2007a}
\bibinfo{author}{\bibfnamefont{W.}~\bibnamefont{Zhang}},
  \bibinfo{author}{\bibfnamefont{H.-S.} \bibnamefont{Chiang}},
  \bibinfo{author}{\bibfnamefont{M.~A.} \bibnamefont{Zudov}},
  \bibinfo{author}{\bibfnamefont{L.~N.} \bibnamefont{Pfeiffer}},
  \bibnamefont{and} \bibinfo{author}{\bibfnamefont{K.~W.} \bibnamefont{West}},
  {\bibinfo{title}{Magnetotransport in a two-dimensional electron system
  in dc electric fields}}, \bibinfo{journal}{Phys. Rev. B}
  \textbf{\bibinfo{volume}{75}}, \bibinfo{pages}{041304(R)}
  (\bibinfo{year}{2007}).

\bibitem[{\citenamefont{Vavilov et~al.}(2007)\citenamefont{Vavilov, Aleiner,
  and Glazman}}]{vavilov:2007}
\bibinfo{author}{\bibfnamefont{M.~G.} \bibnamefont{Vavilov}},
  \bibinfo{author}{\bibfnamefont{I.~L.} \bibnamefont{Aleiner}},
  \bibnamefont{and} \bibinfo{author}{\bibfnamefont{L.~I.}
  \bibnamefont{Glazman}}, {\bibinfo{title}{Nonlinear resistivity of a
  two-dimensional electron gas in a magnetic field}}, \bibinfo{journal}{Phys.
  Rev. B} \textbf{\bibinfo{volume}{76}}, \bibinfo{pages}{115331}
  (\bibinfo{year}{2007}).

\bibitem[{\citenamefont{Coleridge et~al.}(1996)\citenamefont{Coleridge, Hayne,
  Zawadzki, and Sachrajda}}]{coleridge:1996}
\bibinfo{author}{\bibfnamefont{P.}~\bibnamefont{Coleridge}},
  \bibinfo{author}{\bibfnamefont{M.}~\bibnamefont{Hayne}},
  \bibinfo{author}{\bibfnamefont{P.}~\bibnamefont{Zawadzki}}, \bibnamefont{and}
  \bibinfo{author}{\bibfnamefont{A.}~\bibnamefont{Sachrajda}},
  {\bibinfo{title}{Effective masses in high-mobility 2D electron gas
  structures}}, \bibinfo{journal}{Surf. Sci.} \textbf{\bibinfo{volume}{361}},
  \bibinfo{pages}{560} (\bibinfo{year}{1996}).

\bibitem[{\citenamefont{Tan et~al.}(2005)\citenamefont{Tan, Zhu, Stormer,
  Pfeiffer, Baldwin, and West}}]{tan:2005}
\bibinfo{author}{\bibfnamefont{Y.-W.} \bibnamefont{Tan}},
  \bibinfo{author}{\bibfnamefont{J.}~\bibnamefont{Zhu}},
  \bibinfo{author}{\bibfnamefont{H.~L.} \bibnamefont{Stormer}},
  \bibinfo{author}{\bibfnamefont{L.~N.} \bibnamefont{Pfeiffer}},
  \bibinfo{author}{\bibfnamefont{K.~W.} \bibnamefont{Baldwin}},
  \bibnamefont{and} \bibinfo{author}{\bibfnamefont{K.~W.} \bibnamefont{West}},
  {\bibinfo{title}{Measurements of the Density-Dependent Many-Body
  Electron Mass in Two Dimensional GaAs/AlGaAs Heterostructures}},
  \bibinfo{journal}{Phys. Rev. Lett.} \textbf{\bibinfo{volume}{94}},
  \bibinfo{pages}{016405} (\bibinfo{year}{2005}).

\bibitem[{\citenamefont{Hatke et~al.}(2013)\citenamefont{Hatke, Zudov, Watson,
  Manfra, Pfeiffer, and West}}]{hatke:2013}
\bibinfo{author}{\bibfnamefont{A.~T.} \bibnamefont{Hatke}},
  \bibinfo{author}{\bibfnamefont{M.~A.} \bibnamefont{Zudov}},
  \bibinfo{author}{\bibfnamefont{J.~D.} \bibnamefont{Watson}},
  \bibinfo{author}{\bibfnamefont{M.~J.} \bibnamefont{Manfra}},
  \bibinfo{author}{\bibfnamefont{L.~N.} \bibnamefont{Pfeiffer}},
  \bibnamefont{and} \bibinfo{author}{\bibfnamefont{K.~W.} \bibnamefont{West}},
  {\bibinfo{title}{Evidence for effective mass reduction in GaAs/AlGaAs
  quantum wells}}, \bibinfo{journal}{Phys. Rev. B}
  \textbf{\bibinfo{volume}{87}}, \bibinfo{pages}{161307(R)}
  (\bibinfo{year}{2013}).

\bibitem[{\citenamefont{Shchepetilnikov
  et~al.}(2017)\citenamefont{Shchepetilnikov, Frolov, Nefyodov, Kukushkin, and
  Schmult}}]{shchepetilnikov:2017}
\bibinfo{author}{\bibfnamefont{A.~V.} \bibnamefont{Shchepetilnikov}},
  \bibinfo{author}{\bibfnamefont{D.~D.} \bibnamefont{Frolov}},
  \bibinfo{author}{\bibfnamefont{Y.~A.} \bibnamefont{Nefyodov}},
  \bibinfo{author}{\bibfnamefont{I.~V.} \bibnamefont{Kukushkin}},
  \bibnamefont{and} \bibinfo{author}{\bibfnamefont{S.}~\bibnamefont{Schmult}},
  {\bibinfo{title}{Renormalization of the effective mass deduced from the
  period of microwave-induced resistance oscillations in GaAs/AlGaAs
  heterostructures}}, \bibinfo{journal}{Phys. Rev. B}
  \textbf{\bibinfo{volume}{95}}, \bibinfo{pages}{161305(R)}
  (\bibinfo{year}{2017}).

\bibitem[{\citenamefont{Fu et~al.}(2017)\citenamefont{Fu, Ebner, Shi, Zudov,
  Qian, Watson, and Manfra}}]{fu:2017}
\bibinfo{author}{\bibfnamefont{X.}~\bibnamefont{Fu}},
  \bibinfo{author}{\bibfnamefont{Q.~A.} \bibnamefont{Ebner}},
  \bibinfo{author}{\bibfnamefont{Q.}~\bibnamefont{Shi}},
  \bibinfo{author}{\bibfnamefont{M.~A.} \bibnamefont{Zudov}},
  \bibinfo{author}{\bibfnamefont{Q.}~\bibnamefont{Qian}},
  \bibinfo{author}{\bibfnamefont{J.~D.} \bibnamefont{Watson}},
  \bibnamefont{and} \bibinfo{author}{\bibfnamefont{M.~J.}
  \bibnamefont{Manfra}}, {\bibinfo{title}{Microwave-induced resistance
  oscillations in a back-gated GaAs quantum well}}, \bibinfo{journal}{Phys.
  Rev. B} \textbf{\bibinfo{volume}{95}}, \bibinfo{pages}{235415}
  (\bibinfo{year}{2017}).

\bibitem[{not({\natexlab{h}})}]{note:symbols}
\bibinfo{note}{The resistivities represented by $\blacktriangle,\vartriangle$
  ($\blacktriangledown,\triangledown$) were obtained from $\rxx$ and $\ryy$
  \citep{simon:1999} ($\rxx$ and theoretical resistivity product
  \cite{macdonald:2000,oppen:2000}); see \rref{sammon:2019} for details.}

\bibitem[{not({\natexlab{i}})}]{note:17}
\bibinfo{note}{The analysis of the anisotropy ratio in this sample
  \citep{sammon:2019} with theoretical value of $\alpha = 18$ leads to $\gamma
  \approx 0.15$ (see also \rref{note:gamma}) and a ratio of concentrations of
  Coulomb impurities in the spacer $N_1$ to that in the quantum well $N_2$,
  $N_1/N_2 \simeq 60$, much larger than $N_1/N_2 \simeq 10$ estimated in
  \rref{sammon:2018}. Our present experiments, however, suggest lower value of
  $\alpha$, leading to larger $\gamma$ and restoring the agreement with
  \rref{sammon:2018}. Larger $\gamma$ can also result from interface roughness
  scattering, which was not theoretically considered in \rref{sammon:2018}.}

\bibitem[{\citenamefont{Qian et~al.}(2017)\citenamefont{Qian, Nakamura,
  Fallahi, Gardner, and Manfra}}]{qian:2017}
\bibinfo{author}{\bibfnamefont{Q.}~\bibnamefont{Qian}},
  \bibinfo{author}{\bibfnamefont{J.}~\bibnamefont{Nakamura}},
  \bibinfo{author}{\bibfnamefont{S.}~\bibnamefont{Fallahi}},
  \bibinfo{author}{\bibfnamefont{G.~C.} \bibnamefont{Gardner}},
  \bibnamefont{and} \bibinfo{author}{\bibfnamefont{M.~J.}
  \bibnamefont{Manfra}}, {\bibinfo{title}{Possible nematic to smectic
  phase transition in a two-dimensional electron gas at half-filling}},
  \bibinfo{journal}{Nature Commun.} \textbf{\bibinfo{volume}{8}},
  \bibinfo{pages}{1536} (\bibinfo{year}{2017}).

\bibitem[{\citenamefont{Fu et~al.}(2020)\citenamefont{Fu, Shi, Zudov, Gardner,
  Watson, Manfra, Baldwin, Pfeiffer, and West}}]{fu:2020a}
\bibinfo{author}{\bibfnamefont{X.}~\bibnamefont{Fu}},
  \bibinfo{author}{\bibfnamefont{Q.}~\bibnamefont{Shi}},
  \bibinfo{author}{\bibfnamefont{M.~A.} \bibnamefont{Zudov}},
  \bibinfo{author}{\bibfnamefont{G.~C.} \bibnamefont{Gardner}},
  \bibinfo{author}{\bibfnamefont{J.~D.} \bibnamefont{Watson}},
  \bibinfo{author}{\bibfnamefont{M.~J.} \bibnamefont{Manfra}},
  \bibinfo{author}{\bibfnamefont{K.~W.} \bibnamefont{Baldwin}},
  \bibinfo{author}{\bibfnamefont{L.~N.} \bibnamefont{Pfeiffer}},
  \bibnamefont{and} \bibinfo{author}{\bibfnamefont{K.~W.} \bibnamefont{West}},
  {\bibinfo{title}{Anomalous Nematic States in High Half-filled Landau
  Levels}}, \bibinfo{journal}{Phys. Rev. Lett.} \textbf{\bibinfo{volume}{124}},
  \bibinfo{pages}{067601} (\bibinfo{year}{2020}).

\bibitem[{not({\natexlab{j}})}]{note:dis}
\bibinfo{note}{With $i,j = x,y\ (i \neq j)$, the resistivity in units of
  $h/e^2$ is $\tilde\rho_{ii} \simeq \tilde \sigma_{jj}\nu^{-2}$, where $\tilde
  \sigma_{jj} = L_j/2L_i$ and $L_i$ ($L_j$) is the mean free path along the $i$
  ($j$) direction. Without dislocations, $L_y \propto \nu^{-1}$, $L_x \propto
  \nu$, and $\tilde\rho_{xx} \propto \nu^{-4}$, while $\tilde\rho_{yy} \propto
  \nu^0$. With dislocations, $L_y \propto L_x \propto \nu$, i.e., $\tilde
  \sigma_{jj} \propto \nu^0$, and $\tilde\rho_{xx} \propto \tilde\rho_{yy}
  \propto \nu^{-2}$.}

\bibitem[{not({\natexlab{k}})}]{note:dis1}
\bibinfo{note}{Eqs (7) and (8) can be obtained via replacement of hopping time
  of electrons between neighboring stripes $2\tau_B$ by $L_{\rm d}/v$, where
  $v$ is the drift velocity of electron along the stripe.}

\bibitem[{not({\natexlab{l}})}]{note:b}
\bibinfo{note}{The approximate condition for the hQHS phase detection is
  $\tilde\sigma_0 > \alpha^2 (\ttr/\tq) \max\{\ttr/\tq, 11/2\}$.}

\bibitem[{\citenamefont{Borzi et~al.}(2007)\citenamefont{Borzi, Grigera,
  Farrell, Perry, Lister, Lee, Tennant, Maeno, and Mackenzie}}]{borzi:2007}
\bibinfo{author}{\bibfnamefont{R.~A.} \bibnamefont{Borzi}},
  \bibinfo{author}{\bibfnamefont{S.~A.} \bibnamefont{Grigera}},
  \bibinfo{author}{\bibfnamefont{J.}~\bibnamefont{Farrell}},
  \bibinfo{author}{\bibfnamefont{R.~S.} \bibnamefont{Perry}},
  \bibinfo{author}{\bibfnamefont{S.~J.~S.} \bibnamefont{Lister}},
  \bibinfo{author}{\bibfnamefont{S.~L.} \bibnamefont{Lee}},
  \bibinfo{author}{\bibfnamefont{D.~A.} \bibnamefont{Tennant}},
  \bibinfo{author}{\bibfnamefont{Y.}~\bibnamefont{Maeno}}, \bibnamefont{and}
  \bibinfo{author}{\bibfnamefont{A.~P.} \bibnamefont{Mackenzie}},
  {\bibinfo{title}{Formation of a Nematic Fluid at High Fields in
  Sr$_3$Ru$_2$O$_7$}}, \bibinfo{journal}{Science}
  \textbf{\bibinfo{volume}{315}}, \bibinfo{pages}{214} (\bibinfo{year}{2007}).

\bibitem[{\citenamefont{Daou et~al.}(2010)\citenamefont{Daou, Chang, LeBoeuf,
  Cyr-Choiniere, Laliberte, Doiron-Leyraud, Ramshaw, Liang, Bonn, Hardy
  et~al.}}]{daou:2010}
\bibinfo{author}{\bibfnamefont{R.}~\bibnamefont{Daou}},
  \bibinfo{author}{\bibfnamefont{J.}~\bibnamefont{Chang}},
  \bibinfo{author}{\bibfnamefont{D.}~\bibnamefont{LeBoeuf}},
  \bibinfo{author}{\bibfnamefont{O.}~\bibnamefont{Cyr-Choiniere}},
  \bibinfo{author}{\bibfnamefont{F.}~\bibnamefont{Laliberte}},
  \bibinfo{author}{\bibfnamefont{N.}~\bibnamefont{Doiron-Leyraud}},
  \bibinfo{author}{\bibfnamefont{B.~J.} \bibnamefont{Ramshaw}},
  \bibinfo{author}{\bibfnamefont{R.}~\bibnamefont{Liang}},
  \bibinfo{author}{\bibfnamefont{D.~A.} \bibnamefont{Bonn}},
  \bibinfo{author}{\bibfnamefont{W.~N.} \bibnamefont{Hardy}},
  \bibnamefont{et~al.}, {\bibinfo{title}{Broken rotational symmetry in the
  pseudogap phase of a high-Tc superconductor}}, \bibinfo{journal}{Nature}
  \textbf{\bibinfo{volume}{463}}, \bibinfo{pages}{519} (\bibinfo{year}{2010}).

\bibitem[{\citenamefont{Chu et~al.}(2010)\citenamefont{Chu, Analytis, De~Greve,
  McMahon, Islam, Yamamoto, and Fisher}}]{chu:2010}
\bibinfo{author}{\bibfnamefont{J.-H.} \bibnamefont{Chu}},
  \bibinfo{author}{\bibfnamefont{J.~G.} \bibnamefont{Analytis}},
  \bibinfo{author}{\bibfnamefont{K.}~\bibnamefont{De~Greve}},
  \bibinfo{author}{\bibfnamefont{P.~L.} \bibnamefont{McMahon}},
  \bibinfo{author}{\bibfnamefont{Z.}~\bibnamefont{Islam}},
  \bibinfo{author}{\bibfnamefont{Y.}~\bibnamefont{Yamamoto}}, \bibnamefont{and}
  \bibinfo{author}{\bibfnamefont{I.~R.} \bibnamefont{Fisher}},
  {\bibinfo{title}{In-Plane Resistivity Anisotropy in an Underdoped Iron
  Arsenide Superconductor}}, \bibinfo{journal}{Science}
  \textbf{\bibinfo{volume}{329}}, \bibinfo{pages}{824} (\bibinfo{year}{2010}).

\bibitem[{\citenamefont{Okazaki et~al.}(2011)\citenamefont{Okazaki, Shibauchi,
  Shi, Haga, Matsuda, Yamamoto, Onuki, Ikeda, and Matsuda}}]{okazaki:2011}
\bibinfo{author}{\bibfnamefont{R.}~\bibnamefont{Okazaki}},
  \bibinfo{author}{\bibfnamefont{T.}~\bibnamefont{Shibauchi}},
  \bibinfo{author}{\bibfnamefont{H.~J.} \bibnamefont{Shi}},
  \bibinfo{author}{\bibfnamefont{Y.}~\bibnamefont{Haga}},
  \bibinfo{author}{\bibfnamefont{T.~D.} \bibnamefont{Matsuda}},
  \bibinfo{author}{\bibfnamefont{E.}~\bibnamefont{Yamamoto}},
  \bibinfo{author}{\bibfnamefont{Y.}~\bibnamefont{Onuki}},
  \bibinfo{author}{\bibfnamefont{H.}~\bibnamefont{Ikeda}}, \bibnamefont{and}
  \bibinfo{author}{\bibfnamefont{Y.}~\bibnamefont{Matsuda}},
  {\bibinfo{title}{Rotational Symmetry Breaking in the Hidden-Order Phase
  of URu$_2$Si$_2$}}, \bibinfo{journal}{Science}
  \textbf{\bibinfo{volume}{331}}, \bibinfo{pages}{439} (\bibinfo{year}{2011}).

\bibitem[{\citenamefont{Jin et~al.}(2020)\citenamefont{Jin, Tao, Li, Xu, Tang,
  Zhu, Liu, Watanabe, Taniguchi, Hone et~al.}}]{jin:2020}
\bibinfo{author}{\bibfnamefont{C.}~\bibnamefont{Jin}},
  \bibinfo{author}{\bibfnamefont{Z.}~\bibnamefont{Tao}},
  \bibinfo{author}{\bibfnamefont{T.}~\bibnamefont{Li}},
  \bibinfo{author}{\bibfnamefont{Y.}~\bibnamefont{Xu}},
  \bibinfo{author}{\bibfnamefont{Y.}~\bibnamefont{Tang}},
  \bibinfo{author}{\bibfnamefont{J.}~\bibnamefont{Zhu}},
  \bibinfo{author}{\bibfnamefont{S.}~\bibnamefont{Liu}},
  \bibinfo{author}{\bibfnamefont{K.}~\bibnamefont{Watanabe}},
  \bibinfo{author}{\bibfnamefont{T.}~\bibnamefont{Taniguchi}},
  \bibinfo{author}{\bibfnamefont{J.~C.} \bibnamefont{Hone}},
  \bibnamefont{et~al.}, {\bibinfo{title}{Stripe phases in WSe$_2$/WS$_2$
  moir$\acute{e}$ superlattices}}, \bibinfo{journal}{arXiv:2007.12068}.


\bibitem[{\citenamefont{Cao et~al.}(2020)\citenamefont{Cao, Rodan-Legrain,
  Park, Yuan, Watanabe, Taniguchi, Fernandes, Fu, and
  Jarillo-Herrero}}]{cao:2020}
\bibinfo{author}{\bibfnamefont{Y.}~\bibnamefont{Cao}},
  \bibinfo{author}{\bibfnamefont{D.}~\bibnamefont{Rodan-Legrain}},
  \bibinfo{author}{\bibfnamefont{J.~M.} \bibnamefont{Park}},
  \bibinfo{author}{\bibfnamefont{F.~N.} \bibnamefont{Yuan}},
  \bibinfo{author}{\bibfnamefont{K.}~\bibnamefont{Watanabe}},
  \bibinfo{author}{\bibfnamefont{T.}~\bibnamefont{Taniguchi}},
  \bibinfo{author}{\bibfnamefont{R.~M.} \bibnamefont{Fernandes}},
  \bibinfo{author}{\bibfnamefont{L.}~\bibnamefont{Fu}}, \bibnamefont{and}
  \bibinfo{author}{\bibfnamefont{P.}~\bibnamefont{Jarillo-Herrero}},
  {\bibinfo{title}{Nematicity and Competing Orders in Superconducting
  Magic-Angle Graphene}}, \bibinfo{journal}{arXiv:2004.04148}.


\bibitem[{\citenamefont{Fradkin and Kivelson}(1999)}]{fradkin:1999}
\bibinfo{author}{\bibfnamefont{E.}~\bibnamefont{Fradkin}} \bibnamefont{and}
  \bibinfo{author}{\bibfnamefont{S.~A.} \bibnamefont{Kivelson}},
  {\bibinfo{title}{Liquid-crystal phases of quantum Hall systems}},
  \bibinfo{journal}{Phys. Rev. B} \textbf{\bibinfo{volume}{59}},
  \bibinfo{pages}{8065} (\bibinfo{year}{1999}).

\bibitem[{\citenamefont{Zhu et~al.}(2002)\citenamefont{Zhu, Pan, Stormer,
  Pfeiffer, and West}}]{zhu:2002}
\bibinfo{author}{\bibfnamefont{J.}~\bibnamefont{Zhu}},
  \bibinfo{author}{\bibfnamefont{W.}~\bibnamefont{Pan}},
  \bibinfo{author}{\bibfnamefont{H.~L.} \bibnamefont{Stormer}},
  \bibinfo{author}{\bibfnamefont{L.~N.} \bibnamefont{Pfeiffer}},
  \bibnamefont{and} \bibinfo{author}{\bibfnamefont{K.~W.} \bibnamefont{West}},
  {\bibinfo{title}{Density-Induced Interchange of Anisotropy Axes at
  Half-Filled High Landau Levels}}, \bibinfo{journal}{Phys. Rev. Lett.}
  \textbf{\bibinfo{volume}{88}}, \bibinfo{pages}{116803}
  (\bibinfo{year}{2002}).

\bibitem[{\citenamefont{Pan et~al.}(1999)\citenamefont{Pan, Du, Stormer, Tsui,
  Pfeiffer, Baldwin, and West}}]{pan:1999}
\bibinfo{author}{\bibfnamefont{W.}~\bibnamefont{Pan}},
  \bibinfo{author}{\bibfnamefont{R.~R.} \bibnamefont{Du}},
  \bibinfo{author}{\bibfnamefont{H.~L.} \bibnamefont{Stormer}},
  \bibinfo{author}{\bibfnamefont{D.~C.} \bibnamefont{Tsui}},
  \bibinfo{author}{\bibfnamefont{L.~N.} \bibnamefont{Pfeiffer}},
  \bibinfo{author}{\bibfnamefont{K.~W.} \bibnamefont{Baldwin}},
  \bibnamefont{and} \bibinfo{author}{\bibfnamefont{K.~W.} \bibnamefont{West}},
  {\bibinfo{title}{Strongly Anisotropic Electronic Transport at Landau
  Level Filling Factor under a Tilted Magnetic Field}}, \bibinfo{journal}{Phys.
  Rev. Lett.} \textbf{\bibinfo{volume}{83}}, \bibinfo{pages}{820}
  (\bibinfo{year}{1999}).

\bibitem[{\citenamefont{Lilly et~al.}(1999{\natexlab{b}})\citenamefont{Lilly,
  Cooper, Eisenstein, Pfeiffer, and West}}]{lilly:1999b}
\bibinfo{author}{\bibfnamefont{M.~P.} \bibnamefont{Lilly}},
  \bibinfo{author}{\bibfnamefont{K.~B.} \bibnamefont{Cooper}},
  \bibinfo{author}{\bibfnamefont{J.~P.} \bibnamefont{Eisenstein}},
  \bibinfo{author}{\bibfnamefont{L.~N.} \bibnamefont{Pfeiffer}},
  \bibnamefont{and} \bibinfo{author}{\bibfnamefont{K.~W.} \bibnamefont{West}},
  {\bibinfo{title}{Anisotropic States of Two-Dimensional Electron Systems
  in High Landau Levels: Effect of an In-Plane Magnetic Field}},
  \bibinfo{journal}{Phys. Rev. Lett.} \textbf{\bibinfo{volume}{83}},
  \bibinfo{pages}{824} (\bibinfo{year}{1999}{\natexlab{b}}).

\bibitem[{\citenamefont{Jungwirth et~al.}(1999)\citenamefont{Jungwirth,
  MacDonald, Smr\v{c}ka, and Girvin}}]{jungwirth:1999}
\bibinfo{author}{\bibfnamefont{T.}~\bibnamefont{Jungwirth}},
  \bibinfo{author}{\bibfnamefont{A.~H.} \bibnamefont{MacDonald}},
  \bibinfo{author}{\bibfnamefont{L.}~\bibnamefont{Smr\v{c}ka}},
  \bibnamefont{and} \bibinfo{author}{\bibfnamefont{S.~M.}
  \bibnamefont{Girvin}}, {\bibinfo{title}{Field-tilt anisotropy energy in
  quantum Hall stripe states}}, \bibinfo{journal}{Phys. Rev. B}
  \textbf{\bibinfo{volume}{60}}, \bibinfo{pages}{15574} (\bibinfo{year}{1999}).

\bibitem[{\citenamefont{Stanescu et~al.}(2000)\citenamefont{Stanescu, Martin,
  and Phillips}}]{stanescu:2000}
\bibinfo{author}{\bibfnamefont{T.~D.} \bibnamefont{Stanescu}},
  \bibinfo{author}{\bibfnamefont{I.}~\bibnamefont{Martin}}, \bibnamefont{and}
  \bibinfo{author}{\bibfnamefont{P.}~\bibnamefont{Phillips}},
  {\bibinfo{title}{Finite-Temperature Density Instability at High Landau
  Level Occupancy}}, \bibinfo{journal}{Phys. Rev. Lett.}
  \textbf{\bibinfo{volume}{84}}, \bibinfo{pages}{1288} (\bibinfo{year}{2000}).

\bibitem[{\citenamefont{Zhu et~al.}(2009)\citenamefont{Zhu, Sambandamurthy,
  Engel, Tsui, Pfeiffer, and West}}]{zhu:2009}
\bibinfo{author}{\bibfnamefont{H.}~\bibnamefont{Zhu}},
  \bibinfo{author}{\bibfnamefont{G.}~\bibnamefont{Sambandamurthy}},
  \bibinfo{author}{\bibfnamefont{L.~W.} \bibnamefont{Engel}},
  \bibinfo{author}{\bibfnamefont{D.~C.} \bibnamefont{Tsui}},
  \bibinfo{author}{\bibfnamefont{L.~N.} \bibnamefont{Pfeiffer}},
  \bibnamefont{and} \bibinfo{author}{\bibfnamefont{K.~W.} \bibnamefont{West}},
  {\bibinfo{title}{Pinning Mode Resonances of 2D Electron Stripe Phases:
  Effect of an In-Plane Magnetic Field}}, \bibinfo{journal}{Phys. Rev. Lett.}
  \textbf{\bibinfo{volume}{102}}, \bibinfo{pages}{136804}
  (\bibinfo{year}{2009}).

\bibitem[{\citenamefont{Shi et~al.}(2016{\natexlab{b}})\citenamefont{Shi,
  Zudov, Watson, Gardner, and Manfra}}]{shi:2016c}
\bibinfo{author}{\bibfnamefont{Q.}~\bibnamefont{Shi}},
  \bibinfo{author}{\bibfnamefont{M.~A.} \bibnamefont{Zudov}},
  \bibinfo{author}{\bibfnamefont{J.~D.} \bibnamefont{Watson}},
  \bibinfo{author}{\bibfnamefont{G.~C.} \bibnamefont{Gardner}},
  \bibnamefont{and} \bibinfo{author}{\bibfnamefont{M.~J.}
  \bibnamefont{Manfra}}, {\bibinfo{title}{Evidence for a new symmetry
  breaking mechanism reorienting quantum Hall nematics}},
  \bibinfo{journal}{Phys. Rev. B} \textbf{\bibinfo{volume}{93}},
  \bibinfo{pages}{121411(R)} (\bibinfo{year}{2016}{\natexlab{b}}).

\bibitem[{\citenamefont{Shi et~al.}(2017{\natexlab{b}})\citenamefont{Shi,
  Zudov, Qian, Watson, and Manfra}}]{shi:2017c}
\bibinfo{author}{\bibfnamefont{Q.}~\bibnamefont{Shi}},
  \bibinfo{author}{\bibfnamefont{M.~A.} \bibnamefont{Zudov}},
  \bibinfo{author}{\bibfnamefont{Q.}~\bibnamefont{Qian}},
  \bibinfo{author}{\bibfnamefont{J.~D.} \bibnamefont{Watson}},
  \bibnamefont{and} \bibinfo{author}{\bibfnamefont{M.~J.}
  \bibnamefont{Manfra}}, {\bibinfo{title}{Effect of density on quantum
  Hall stripe orientation in tilted magnetic fields}}, \bibinfo{journal}{Phys.
  Rev. B} \textbf{\bibinfo{volume}{95}}, \bibinfo{pages}{161303(R)}
  (\bibinfo{year}{2017}{\natexlab{b}}).

\bibitem[{\citenamefont{Falson et~al.}(2018)\citenamefont{Falson, Tabrea,
  Zhang, Sodemann, Kozuka, Tsukazaki, Kawasaki, von Klitzing, and
  Smet}}]{falson:2018}
\bibinfo{author}{\bibfnamefont{J.}~\bibnamefont{Falson}},
  \bibinfo{author}{\bibfnamefont{D.}~\bibnamefont{Tabrea}},
  \bibinfo{author}{\bibfnamefont{D.}~\bibnamefont{Zhang}},
  \bibinfo{author}{\bibfnamefont{I.}~\bibnamefont{Sodemann}},
  \bibinfo{author}{\bibfnamefont{Y.}~\bibnamefont{Kozuka}},
  \bibinfo{author}{\bibfnamefont{A.}~\bibnamefont{Tsukazaki}},
  \bibinfo{author}{\bibfnamefont{M.}~\bibnamefont{Kawasaki}},
  \bibinfo{author}{\bibfnamefont{K.}~\bibnamefont{von Klitzing}},
  \bibnamefont{and} \bibinfo{author}{\bibfnamefont{J.~H.} \bibnamefont{Smet}},
  {\bibinfo{title}{A cascade of phase transitions in an orbitally mixed
  half-filled Landau level}}, \bibinfo{journal}{Science Advances}
  \textbf{\bibinfo{volume}{4}} (\bibinfo{year}{2018}).


\bibitem[{\citenamefont{Hossain et~al.}(2018)\citenamefont{Hossain, Mueed, Ma,
  Chung, Pfeiffer, West, Baldwin, and Shayegan}}]{hossain:2018}
\bibinfo{author}{\bibfnamefont{M.~S.} \bibnamefont{Hossain}},
  \bibinfo{author}{\bibfnamefont{M.~A.} \bibnamefont{Mueed}},
  \bibinfo{author}{\bibfnamefont{M.~K.} \bibnamefont{Ma}},
  \bibinfo{author}{\bibfnamefont{Y.~J.} \bibnamefont{Chung}},
  \bibinfo{author}{\bibfnamefont{L.~N.} \bibnamefont{Pfeiffer}},
  \bibinfo{author}{\bibfnamefont{K.~W.} \bibnamefont{West}},
  \bibinfo{author}{\bibfnamefont{K.~W.} \bibnamefont{Baldwin}},
  \bibnamefont{and} \bibinfo{author}{\bibfnamefont{M.}~\bibnamefont{Shayegan}},
  {\bibinfo{title}{Anomalous coupling between magnetic and nematic orders
  in quantum Hall systems}}, \bibinfo{journal}{Phys. Rev. B}
  \textbf{\bibinfo{volume}{98}}, \bibinfo{pages}{081109(R)}
  (\bibinfo{year}{2018}).

\bibitem[{\citenamefont{Simon}(1999)}]{simon:1999}
\bibinfo{author}{\bibfnamefont{S.~H.} \bibnamefont{Simon}},
  {\bibinfo{title}{Comment on ``Evidence for an Anisotropic State of
  Two-Dimensional Electrons in High Landau Levels''}}, \bibinfo{journal}{Phys.
  Rev. Lett.} \textbf{\bibinfo{volume}{83}}, \bibinfo{pages}{4223}
  (\bibinfo{year}{1999}).

\bibitem[{\citenamefont{Sammon et~al.}(2018)\citenamefont{Sammon, Zudov, and
  Shklovskii}}]{sammon:2018}
\bibinfo{author}{\bibfnamefont{M.}~\bibnamefont{Sammon}},
  \bibinfo{author}{\bibfnamefont{M.~A.} \bibnamefont{Zudov}}, \bibnamefont{and}
  \bibinfo{author}{\bibfnamefont{B.~I.} \bibnamefont{Shklovskii}},
  {\bibinfo{title}{Mobility and quantum mobility of modern GaAs/AlGaAs
  heterostructures}}, \bibinfo{journal}{Phys. Rev. Materials}
  \textbf{\bibinfo{volume}{2}}, \bibinfo{pages}{064604} (\bibinfo{year}{2018}).

\end{thebibliography}
\end{document}